\documentclass{aa}  

\usepackage{placeins}
\usepackage{graphicx,subcaption}
\usepackage{lscape}
\usepackage{txfonts}
\usepackage{multicol}
\usepackage{multirow}
\usepackage{hyperref}
\hypersetup{
    colorlinks=true,
    linkcolor=blue,
    filecolor=magenta,      
    urlcolor=blue,
    citecolor=blue
    }
\usepackage{comment}

\defcitealias{SACCHI2025}{S+25}
\defcitealias{AKINS2024}{A+24}
\defcitealias{VERGARA2023}{V+23}
\defcitealias{VERGARA2024}{V+24}
\defcitealias{VERGARA2025a}{V+25a}
\defcitealias{VERGARA2025b}{V+25b}

\begin{document}

   \title{Constraints on dynamically-formed massive black holes in Little Red Dots from X-ray non-detections}

   \author{M. Liempi
          \inst{1}
          \and
            D.R.G. Schleicher\inst{1}
            \and M. A. Latif\inst{2} 
            \and
            R. Schneider\inst{1,3,4,5}
            \and
            F. Flammini Dotti\inst{6,7,1}
             \and
            A. Escala\inst{8}
            \and 
            M.C. Vergara\inst{9}
            }

\institute{
         Dipartimento di Fisica, Sapienza Universit\`a di Roma, Piazzale Aldo Moro 5, 00185 Rome, Italy\\
        \email{gonzalez.liempi@uniroma1.it}
        \and Physics Department, College of Science, United Arab Emirates University, PO Box 15551,Al-Ain, United A\\ \email{latifne@gmail.com}
        \and INFN – Sezione Roma1, Dipartimento di Fisica, Università di Roma La Sapienza, Piazzale Aldo Moro 2, I-00185 Roma, Italy
         \and
         Sapienza School for Advanced Studies, Viale Regina Elena 291, I- 00161 Roma, Italy 
         \and INAF – Osservatorio Astronomico di Roma, via di Frascati 33, 00078 Monte Porzio Catone, Italy
         \and Department of Physics, New York University Abu Dhabi, PO Box 129188 Abu Dhabi, UAE
         \and  Center for Astrophysics and Space Science (CASS), New York University Abu Dhabi, PO Box 129188, Abu Dhabi, UAE
        \and
        Departamento de Astronom\'ia, Universidad de Chile, Casilla 36-D, Santiago, Chile
        \and
        Astronomisches Rechen-Institut, Zentrum für Astronomie, University of Heidelberg, Mönchhofstrasse 12-14, 69120, Heidelberg, Germany 
         }

   \date{Received XXXX; accepted XXXX}

  \abstract
   { The existence of extremely massive and compact galaxies (Little Red Dots or LRDs) at $z\gtrsim 2$ challenges models of early structure formation, suggesting extremely rapid stellar and black hole (BH) assembly. These galaxies are efficient
   environments for rapid BH growth, yet several LRDs show no evidence of strong emission from the active galactic nuclei (AGN) in X-rays.  Our work utilizes a subsample of X-ray non-detected LRDs to check whether the collision-based BH formation scenario is compatible with the non-detections and constrain both physical parameters (e.g., metallicity) and observational parameters (e.g., column density). Our results show that LRDs might be ideal birthplaces to form massive BHs, particularly in the case of a mass-radius relation $R_{\rm gal}\propto M^{0.6}_{\rm gal}$ (similar to spiral galaxies in the Local Universe). 
   Given the large stellar densities, collision-based models suggest seed masses greater than those observed in the local Universe and those are compatible with the mass-radius relation of high-redshift BHs. 
   We model the BH seed formation and subsequent X-ray emission investigating both the physical and observational parameter space to compare with the observed X-ray upper limits in the soft band ($0.3-2$\,keV) and the hard band ($2-7$\,keV). 
   We find exponents in the mass-radius relation above  $0.55$ favor the collision-based scenario; however, consistency with the stacked X-ray analysis requires specific combinations of accretion and obscuration parameters. Additionally, we find that constant or increasing star formation rate scenarios assuming high Eddington ratios and large duty cycles are feasible but require larger column densities and/or a higher level of metal enrichment in the shielding columns. Alternatively, moderate sub-Eddington accretion rates seem to be sufficient to reconcile the massive seeds with their final observed masses, consistent with the observed X-ray weakness. Overall, we conclude that even in cases where LRDs were initially starburst galaxies, they should evolve into an AGN.}

   \keywords{galaxies: evolution --- galaxies: formation--- galaxies: nuclei --- galaxies: high-redshift
---quasars: supermassive black holes --- X-rays: galaxies }
 \authorrunning{Liempi et. al}
   \maketitle

\section{Introduction}

The detection of extremely compact and massive galaxies at high-redshift by JWST challenges the current knowledge of galaxy formation and evolution. These galaxies, the so-called Little Red Dots (LRDs), have masses ranging between $10^8$ and $10^{12}$\,M$_{\odot}$ in an effective radius of $3 - 300$\,pc \citep[e.g.,][]{MATTHEE2023, GREENE2024, AKINS2024}. What characterizes an LRD is its `V-shaped' Spectral Energy Distribution (SED), which features an inclined red continuum in the optical rest frame together with a blue UV counterpart in the rest frame \citep{KOCEVSKI2023, AKINS2023, MAIOLINO2024}. The large sensitivity of JWST surveys, particularly the JWST Advanced Deep Extragalactic Survey (JADES) \citep{EISENSTEIN2023a, EISENSTEIN2023B}, facilitates the discovery and characterization of LRDs over the GOODS-South and GOODS-North fields, contributing to the measurement of their properties \citep[e.g.,][]{RINALDI2025}. However, the origin of this peculiar spectrum is still strongly debated. The slopes in the optical band are consistent with (i) emission from dusty star-formation \citep[i.e., emission dominated by young stellar populations;][]{WILLIAMS2024,PEREZGONZALEZ2024} or (ii) reddened Active Galactic Nuclei (AGN) where thermal emission from the accretion disk dominates on scales less than $1$\,pc with temperatures higher than $10^5$\,K \citep[][]{LABBE2025,MATTHEE2023}. Similarly, the excess of UV emission might be explained by light from the central AGN and stellar emission emerging from a relatively dust-free host, or by light escaping without attenuation due to patchy dust, as observed in some lower redshift dusty star-forming galaxies \citep[e.g.,][]{CASEY2014} and red quasars \citep{GLIKMAN2023}.

While the broad emission lines (FWHM $\sim 2000\,{\rm km\,s}^{-1}$) observed in many LRDs suggest the presence of accreting supermassive black holes (SMBHs) with masses of at least $10^7-10^8$\,M$_\odot$ \citep{MATTHEE2023}, confirming this AGN nature in the X-ray regime has proven difficult. Recent stacking analyses of X-ray non-detected sources have yielded strict upper limits. For instance, \citet{SACCHI2025} (hereafter \citetalias{SACCHI2025}) constrained the emission of galaxies lacking individual X-ray detections. They found upper limits that are incompatible with unobscured super-Eddington accretion models. Their results imply that if massive BHs are present, they must be either significantly less massive than optical estimates suggest, or are buried under extreme column densities ($N_{\rm H} \gtrsim 10^{25}$\,cm$^{-2}$). This characteristic ``X-ray weakness'' is supported by independent studies. \cite{ANANNA2024} found no significant detection in a stack of 21 high-redshift LRDs, placing upper limits on BH masses of $\lesssim(1.5-16)\times 10^6$\,M$_\odot$ assuming Eddington-limited accretion. Similarly, \cite{YUE2024} reported only tentative detections in a stack of 34 LRDs, suggesting that any central engine must be intrinsically weak or heavily obscured. This discrepancy between the optical evidence for massive BHs and the general lack of strong X-ray emission presents a major open question regarding the nature and growth of LRDs.

Despite these observational puzzles, the structural properties of LRDs make them unique laboratories for BH formation. On average, LRDs have large stellar masses ($M_{\rm gal}$) of a few $10^{10}$\,M$_\odot$ within radii of $R_{\rm gal}\sim 135$\,pc at $z\sim 7-9$ \citep{BAGGEN2023}, with some systems being even more compact ($R_{\rm gal} < 35$\,pc, \citealp{FURTAK2023}). The mean density in the core of LRDs is $\sim 10^{4}$\,M$_\odot$\,pc$^{-3}$, with the highest core densities reaching $10^{8}$\,M$_\odot$\,pc$^{-3}$ \citep{GUIA2024}. Theoretical models predict that in environments exceeding $10^{7}$\,M$_\odot$\,pc$^{-3}$, runaway stellar collisions become inevitable. Numerical simulations show that stellar mergers can produce a single supermassive star, which then collapses directly into an intermediate-mass BH \citep{FUJIII2024}. Other pathways include the growth of a massive star within a dense cluster formed from a fragmented gas cloud \citep{TAGAWA2020}. Recent N-body models further support this scenario, showing that collisions in dense clusters can efficiently produce intermediate-mass BHs \citep[][ hereafter V+]{VERGARA2023,VERGARA2024,VERGARA2025a,VERGARA2025b}. Indeed, semi-analytic models of Nuclear Star Clusters (NSCs) suggest that this collision-based channel likely makes a relevant contribution to the total population of SMBHs \citep{LIEMPI2025}.

In this paper, we aim to constrain the properties of massive BHs formed via stellar collisions in LRDs by exploiting the reported limits on their X-ray emission. We construct a dynamical evolution model for LRDs to predict the mass of collisionally formed seeds and their subsequent X-ray signatures. By comparing these predictions with the stacking limits from \citetalias{SACCHI2025}, we identify the regions of parameter space---specifically regarding the galaxy mass-radius relation and BH accretion duty cycles---that are consistent with the current non-detections.

This work is organized in the following sections: in Section \ref{sec:data} we briefly describe the data collection of our sample. The model for the temporal evolution of LRDs is described in Section \ref{sec:dynamics}. We describe the obtained results in Section~\ref{sec:results} and discuss them in Section~\ref{discussion}.

\section{Data and general framework} \label{sec:Models}
In this Section, we introduce the observational data employed for our data analysis. Subsequently, we introduce a simplified dynamical model for the evolution of LRDs. We aim to estimate the masses of SMBHs that could form in these dense systems. Finally, we discuss the observational constraints from the X-ray background and our emission model to estimate the expected X-ray emission from the BH population.

\subsection{Population data} \label{sec:data}
In this work, we use a subsample of 55 galaxies selected by \citetalias{SACCHI2025} from the parent sample of \cite{KOCEVSKI2025}. We here consider the galaxies for which deep Chandra data have been available without individual X-ray detections. The authors employed the stacking technique to combine the data from the different sources and derive an upper limit on the X-ray flux that may come from this population. The parent sample contains the photometric redshift and the absolute magnitude in the UV band (${\rm M}_{\rm UV}$) of 341 LRDs spanning the redshift range $z\sim 2-11$ using data from the CEERS, PRIMER, JADES, UNCOVER and NGDEEP surveys. In contrast, the subsample of \citetalias{SACCHI2025} contains galaxies in the JADES and NGDEEP surveys, which do not have X-ray detections in a similar redshift range $z\sim 3-11$.

\begin{figure}[!h]
    \centering
    \includegraphics[scale=0.9]{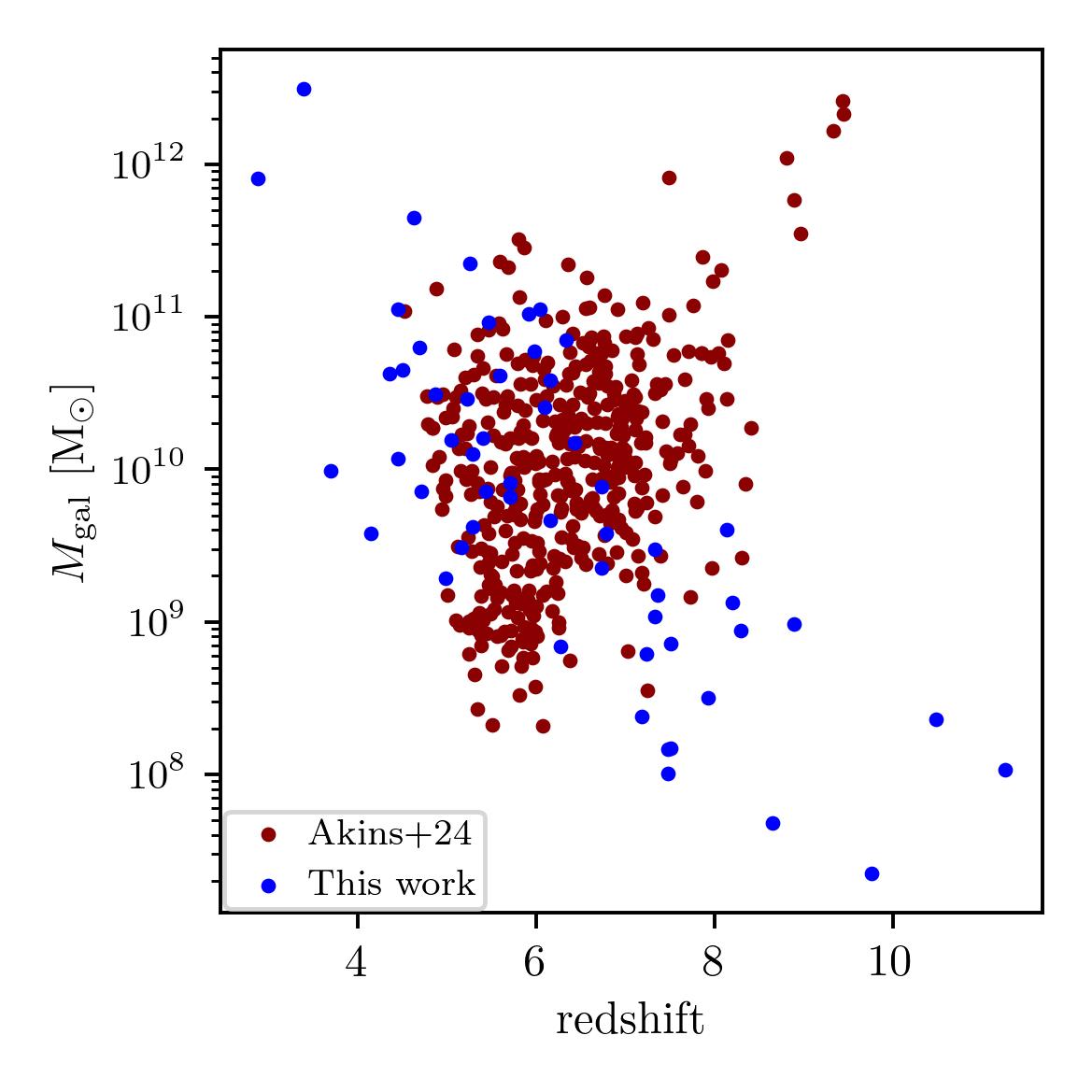}
    \caption{Galaxy stellar masses obtained for the subsample of \citetalias{SACCHI2025} using the previous observed scaling correlation between ${\rm M}_{\rm UV}$ and $M_{\rm gal}$ (Eq. \ref{eq:stellarMassCorrelation}) compared to the stellar masses of the galaxies in the sample of \citetalias{AKINS2024}, both shown as a function of redshift. }
    \label{fig:redshiftMassStellar}
\end{figure}

As input for our dynamical model, we estimate the stellar mass ($M_{\rm gal}$) for our galaxies using the $M_{\rm UV}-M_{\rm gal}$ correlation \citep[e.g.,][]{GONZALEZ2011,DUNCAN2014} at different redshifts.  Specifically, it is well known that $M_{\rm gal}$ is correlated with the UV absolute magnitude $M_{\rm UV}$ as
\begin{equation}
    \log{M_{\rm gal}}  = a\cdot M_{\rm UV} + b,\label{eq:stellarMassCorrelation}
\end{equation}
where $a,b$ are free parameters that depend on the redshift.  In Table \ref{tab:stellarMassParameters} we list the values adopted for the free parameters in Eq. \ref{eq:stellarMassCorrelation}. In Fig. \ref{fig:redshiftMassStellar}, we show the comparison between the resulting stellar masses and the masses reported in the sample of \cite{AKINS2024} (hereafter \citetalias{AKINS2024}). We find a good agreement regarding the overall mass range, while we note that the redshift distribution in the subsample of \citetalias{SACCHI2025} is slightly different, extending both to somewhat lower ($z\sim 2.8$) and somewhat higher ($z\sim 11$) redshifts compared to the \citetalias{AKINS2024} sample.

\begin{table}
\centering
\caption{Summary of the parameters used in the scaling laws (Eq.~\ref{eq:stellarMassCorrelation}) to estimate the stellar mass of the galaxies.}
\begin{tabular}{ccccc}
\hline
$z$ & $a$     & $b$       & Intrinsic & Reference          \\
(1) & (2) & (3)     & (4)       & (5)                \\ \hline
3 & $-0.41^{+0.10}_{-0.10}$ & $+1.200^{+0.2}_{-0.2}$ & 0.30 dex  & \cite{LEE2012}     \\
4 & $-0.54^{+0.07}_{-0.07}$ & $-1.700^{+0.3}_{-0.3}$ & 0.53 dex  & \cite{SONG2016}    \\
5 & $-0.50^{+0.07}_{-0.07}$ & $-0.900^{+0.3}_{-0.3}$ & 0.42 dex  & \cite{SONG2016}    \\
6 & $-0.57^{+0.05}_{-0.05}$ & $-1.064^{+0.2}_{-0.2}$ & 0.30 dex  & \cite{STEFANON2021}\\
7 & $-0.49^{+0.05}_{-0.05}$ & $-1.056^{+0.2}_{-0.2}$ & 0.30 dex  & \cite{STEFANON2021}\\
8 & $-0.49^{+0.05}_{-0.05}$ & $-1.047^{+0.2}_{-0.2}$ & 0.30 dex  & \cite{STEFANON2021}\\
9 & $-0.46^{+0.05}_{-0.05}$ & $-1.028^{+0.2}_{-0.2}$ & 0.30 dex  & \cite{STEFANON2021}\\ 
10\tablefootmark{a}& $-0.46^{+0.05}_{-0.05}$ & $-1.023^{+0.2}_{-0.2}$ & 0.30 dex  & \cite{STEFANON2021}\\ \hline
\end{tabular}
\tablefoot{(1) Redshift range $z$; (2) slope of the correlation $a$; (3) intercept coefficient $b$; (4) intrinsic scatter; (5) Reference for the values adopted here.  \tablefoottext{a} {As there is no published correlation for redshift 11 we use the same as for $z=10$ in \citet{STEFANON2021}.}} \label{tab:stellarMassParameters}
\end{table}

\begin{figure}[!h]
    \centering
\includegraphics[height=0.40\textwidth,
  keepaspectratio]{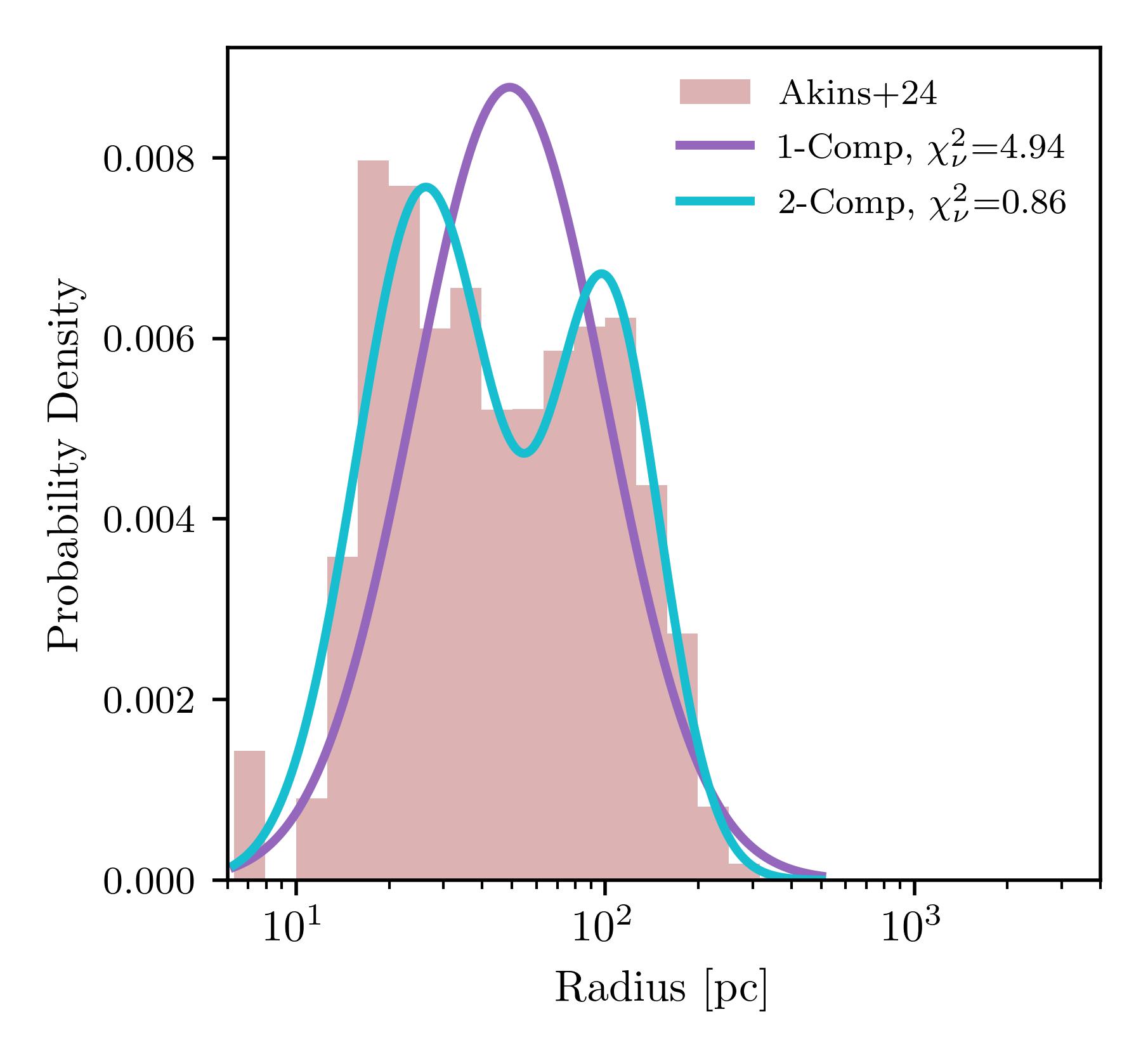}
    \caption{Probability density distribution of effective radii ($R_{\text{gal}}$) for the galaxy sample from \citetalias{AKINS2024}. The shaded red histogram shows the observed data binned in logarithmic intervals. The solid curves represent the best-fit Gaussian mixture models, assuming a single component (purple) and two components (cyan). We show the reduced chi-squared ($\chi^2_\nu$) for each model.}
    \label{fig:radiusDistribution}
\end{figure}

As the radii for the sources found in \citetalias{SACCHI2025} sample have not been yet published, we assume here that they follow the same distribution of the sample in \citetalias{AKINS2024}, where it appears roughly independent of mass (see Appendix \ref{Apppendix:A}). We obtain the effective radius of the galaxies in \citetalias{AKINS2024} sample. In there, the authors multiply the radius (in mas) by the angular distance $D_A(z)$. As galaxy sizes span on several orders of magnitude, we performed the fitting in logarithmic space, defined as $x = \log_{10}(R_{\text{gal}}/\text{pc})$. We employed a Gaussian Mixture Model approach, where the distribution is described as the sum of $K$ independent Gaussian components. The probability density $P(x)$ is given by $P(x) = \sum_{k=1}^{K} w_k \frac{1}{\sqrt{2\pi\sigma_k^2}} \exp\left( -\frac{(x - \mu_k)^2}{2\sigma_k^2} \right),$ where $w_k$, $\mu_k$, and $\sigma_k$ are the weight, mean, and standard deviation of the $k$-th component, respectively, subject to the constraint $\sum w_k = 1$. The parameters were estimated using the Expectation-Maximization algorithm. We compared models with $K=1$ and $K=2$ components (see Fig. \ref{fig:radiusDistribution}), evaluating the goodness-of-fit using the reduced chi-squared statistic ($\chi^2_\nu$) calculated over the histogram bins assuming a fixed size equals to  $0.2$\, dex. Prior to fitting, we excluded a single object with an effective radius $R_{\text{eff}} > 1$ kpc. As this object was the sole galaxy in the sample with such a large extent, treating it as part of the continuous distribution would disproportionately affect the parametric fit or require an statistically unjustified third component; removing it allows for a robust characterization of the primary compact and extended populations. Even though the redshift and the mass distributions of the \citetalias{AKINS2024} sample and the \citetalias{SACCHI2025} subsample are slightly different, we assume that the radius distribution used here provides a conservative estimate as discussed in Appendix \ref{sec:appendix_comparison}. We checked how results are affected by sampling from the one and two components distributions finding no relevant changes.

The cosmological parameters adopted for all calculations corresponds to the \citet{PLANCK2020} results, with $H_0=67.66\,{\rm km\, s^{-1}\, Mpc^{-1}}$, $\Omega_{\rm m,0}=0.31$. All magnitudes are in the AB system \citep{OKE1974}, as in \citetalias{AKINS2024}.

\subsection{Dynamics of Little Red Dots} \label{sec:dynamics}

In this subsection we describe our prescription for the mass and size evolution of LRDs.

We employ a simplified toy model for the evolution of the mass and the radius of the LRD in which  both the parameters $M_{\rm gal}$ and $R_{\rm gal}$ are approximated as power laws of time. Specifically, the evolution of the mass is given by
\begin{equation}
    M_{\rm gal} (t) \propto t^\alpha \label{eq:mass},
\end{equation}
with $\alpha=[0.1, 3]$ the exponent of the power law. On the other hand, the evolution of the radius depends implicitly on time. It assumes a correlation between the stellar mass and the radius, i.e.,
\begin{equation}
    R_{\rm gal} (t) \propto M^{\beta}_{\rm gal}(t) \label{eq:radius},
\end{equation}
where $\beta$ is the exponent of the correlation $R_{\rm gal}-M_{\rm gal}$\footnote{Note that the normalization of the correlation is individual for each source in order to match the final radius sampled from the distribution.} and assumed to be in the range of $0.1-1.5$. This parameter is a key point of observational debate at high-redshift. For example, \cite{ALLEN2025} measured a constant size-mass relation slope of $\beta \approx 0.215$ for star-forming galaxies from $z=3-9$. In contrast, \cite{ORMEROD2024} found that this relationship breaks down at $z > 3$, with galaxy sizes showing little to no dependence on stellar mass (implying $\beta \approx 0$). For NSCs, we have $\beta\sim0.5$, while  early-type galaxies have typical values
of $\beta = 0.55 - 0.6$ \citep{SHEN2003} and late-type galaxies have $\beta = 0.22 - 0.27$ \citep{SHEN2003, LANGE2015}. Our model, which tests a wide range of $\beta$, can therefore provide constraints on which evolutionary track is more consistent with the observed properties of LRDs.

\subsection{Black hole formation via stellar collisions}\label{sec:BHFormation}

The work of \cite{ESCALA2021} provides observational evidence that supports the formation  of a massive BH due to stellar runaway collisions in dense stellar systems. In  systems with collisional timescales ($t_{\rm coll}$) shorter than the age of the system, a global instability leads to massive object formation  via runaway stellar collisions. This scenario has been tested with dedicated $N$-body direct simulations by \citetalias{VERGARA2023} as well as through a detailed comparison and analysis with literature data \citepalias{VERGARA2024,VERGARA2025a,VERGARA2025b}.

\begin{figure}
\centering
\includegraphics{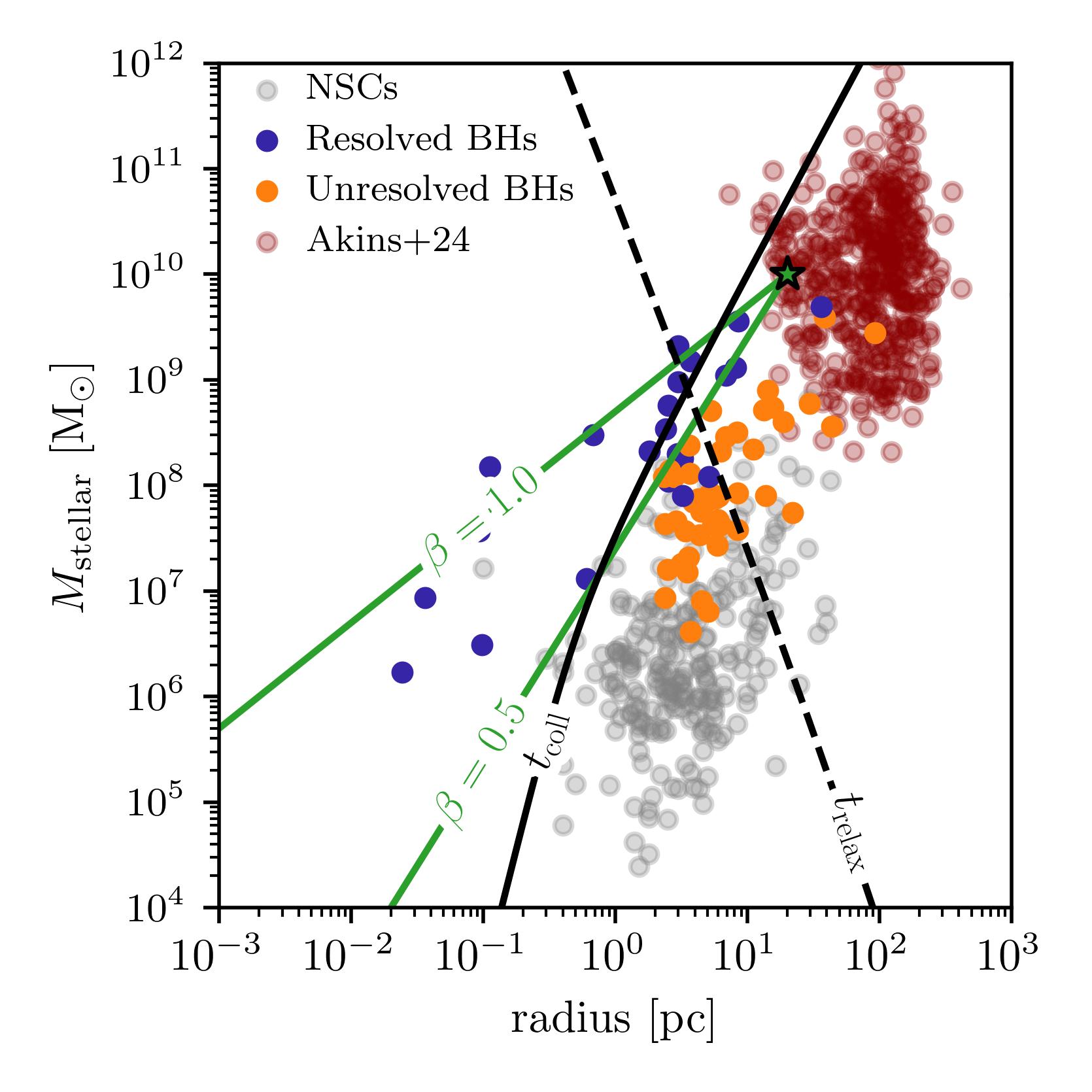}
\caption{Mass-radius diagram for stellar systems to illustrate relevant timescales and the position of different sources in different parts of the parameter space.  Red dots are LRDs from \citetalias{AKINS2024}, gray dots are NSCs in the local universe \citep{GEORGIEV2016,NEUMAYER2020}, blue dots are well resolved BHs, while orange dots are spatially unresolved BHs \citep{GULTEKIN2009,ESCALA2021}. The solid black line represents the collisional timescale $t_{\rm coll} = 13.7$\, Gyr, while the dashed black line is the relaxation time for $t_{\rm relax}=13.7$\, Gyr. We include (in green) a fiducial galaxy with mass $10^{10}$\,M$_\odot$ and radius $20$\,pc to show its trajectory in the mass-radius plane for $\beta=1$ and $0.5$ defined in Eq. \ref{eq:radius}. The time evolution of the green line is from the left to the right. }\label{fig:massRadiusDiagram}
\end{figure}

In Fig. \ref{fig:massRadiusDiagram}, we show a mass-radius diagram divided in two main regions by the condition of the collisional timescale (black solid line, defined in Eq. \ref{eq:collisionTimescale}) being equal to the age of the Universe ($t_{\rm coll}=13.7$~Gyr). On the left side of the collisional timescale, collisions are relevant. In there, the collision timescale is shorter than the age of the system (i.e., $t_{\rm coll} \leq t_{\rm H}$), with $t_{\rm H}$ the age of the system. In the right side of the collisional timescale line, collisions are less relevant (but still happen) as the collision timescale is larger than the age of the system ($t_{\rm H}<t_{\rm coll}$). We show how the observed NSCs reside in a stable regions within the mass-radius diagram of stellar systems, delimited by a line for which the collision timescale $t_{\rm coll}$ becomes $13.7$~Gyr. We define it as \cite{BINNEY2009}
\begin{equation}
    t_{\rm coll} = \sqrt{\frac{R_{\rm gal}}{GM_{\rm gal}(n\Sigma_0)^2}}, \label{eq:collisionTimescale}
\end{equation}
with $n=\eta M_{\rm gal}/ (R_{\rm gal}^3{\rm M}_\odot)$ the number density of stars ($\eta=3/4\pi$ for systems with spherical symmetry), $\Sigma_0=16\sqrt{\pi}(1+\Theta)\,{\rm R}_\odot^2$ the cross section and $\Theta=9.54\left[(M_\star\,{\rm R}_\odot)/(R_\star\, {\rm M}_\odot)\right]\left(100\, {\rm km\,s^{-1}}/\sigma\right)^2$ the Safronov number, and $\sigma=\sqrt{GM_{\rm gal}/R_{\rm gal}}$, which is the velocity dispersion under the assumption of virial equilibrium.

In Fig. \ref{fig:massRadiusDiagram}, we also show the mass and radius (estimated as $=0.5d_{\rm resol}$, with $d_{\rm resol}$  the spatial resolution) of SMBHs in the sample of \cite{GULTEKIN2009}, where we refer to well resolved SMBHs to candidates with influence radius ($R_{\rm inf}$) larger than three spatial resolutions. These well-resolved BHs are located in the regime where stellar collisions are relevant for the dynamics of the system. Spatially unresolved SMBHs (defined as $R_{\rm ind}<3 d_{\rm resol}$) are in the stable area. In those case, the radius should be regarded only as an upper limit. This does not represent an issue as better resolution might shift the positions of the unresolved population to the left. 

Similarly, LRDs are close to the regime where collisions will be relevant, but in a different position on the mass-radius diagram. We include a fiducial galaxy in green (with final mass equals to $10^{10}$\,M$_\odot$ and radius $20$\, pc) to show how it moves in the diagram (from the left to the right) as it evolves according to the power-law model (Eqs. \ref{eq:mass} and \ref{eq:radius}) and reaches a final density $\sim10^7$\,M$_\odot$\,pc$^{-3}$ (assuming a system composed by Solar mass stars) similar to the LRD densities calculated by \cite{GUIA2024}.

As previously mentioned in the introduction, LRDs are potential birthplaces for massive BHs due to the high densities that their cores can reach \citep{GUIA2024}. Different approaches \citep[e.g., analytical, N-Body, and Fokker-Plank models][]{PACUCCI2025} show that the formation of a massive object due to pure stellar dynamics is expected, and most probably formed directly in the supermassive range in the case of LRDs, which were initially even more compact \citep{ESCALA2025}.

In direct N-body simulations investigating this scenario, it was found that the BH formation efficiency, defined as $\epsilon_{\rm BH}= \left(1+M_{\rm stellar}/M_{\rm BH}\right)^{-1}$ (where $M_{\rm stellar}$ is the stellar mass of the system and $M_{\rm BH}$ is the mass of the BH), reached values up to $\sim 50\%$ in dense clusters with stellar masses around $\sim 10^4$\,M$_\odot$ \citepalias{VERGARA2023}. This efficiency represents a saturation limit observed in environments with densities of order $\sim 10^{7}$\,M$_\odot$\,pc$^{-3}$. Such extreme conditions are consistent with the estimated core densities of LRDs \citep{GUIA2024}. Furthermore, comparisons with observations of dense stellar systems suggest that this mechanism can account for BHs with masses up to $\sim 10^7$\,M$_\odot$ \citepalias{VERGARA2024}, supporting the scenario where the dense cores of LRDs are converted efficiently into massive BHs.
They further have shown how this efficiency depends on the ratio of stellar mass to  critical mass $M_{\rm crit}(R)$, which encapsulates the idea of stellar collisions, defined as 
\begin{equation}
    M_{\rm crit}(R) = R^\frac{7}{3}\left(\frac{4\pi M_\star}{3\Sigma_0t_{\rm H}G^\frac{1}{2}} \right)^\frac{2}{3}, 
\end{equation}
where $R$ is the radius of the system, $M_\star$ the mass of a single star, $t_{\rm H}$ the age of the system, and $\Sigma_0$  the effective cross section.

In our model, we employ the fit formula from  \citetalias{VERGARA2025b} to estimate the efficiency of the forming BH, given as\begin{equation}
    \epsilon_{\rm BH} \left( \frac{M_{\rm gal}}{M_{\rm crit}} \right)= \left[ 1+\exp\left(-4.63\left[\log\left(\frac{M_{\rm gal}}{M_{\rm crit}}\right)-4\right]\right)\right]^{-0.1}.\label{eq:blackHoleEfficiency}
\end{equation}

 In galaxies where this condition is reached, we evaluate this efficiency when $t=t_{\rm coll}$, as the value of the efficiency will be largest at that time. Even if the condition is not fulfilled, it is still possible to have efficiencies larger than unity, implying that a relevant fraction of the stellar mass goes into a massive object. We checked that the product $\epsilon_{\rm BH}M_{\rm gal}$ monotonically increases with time in these models. Therefore, we evaluate the efficiency at the observed age of the galaxy when $t=t_{\rm age}$, where we estimate the age of the system based on its observed redshift. In summary, we assign the BH mass as $M_{\rm BH}=\epsilon_{\rm BH}M_{\rm gal}$, where the efficiency is given by Eq. \ref{eq:blackHoleEfficiency}. The ratio between the galaxy mass and the critical mass ($M_{\rm gal}/M_{\rm crit}$) is evaluated at $t=t_{\rm coll}$ or $t=t_{\rm age}$ accordingly. The maximum BH masses for which the models are validated is on the order of $10^{7}$~M$_\odot$, which we take as an upper limit \citepalias{VERGARA2023, VERGARA2024}. While it may be possible that the limit of validity of the model could extend up to higher masses, we nonetheless note that additional effects may become relevant. Stellar evolution is uncertain at these mass scales and for those large dynamical ranges. We adopt this upper limit because the observational samples in the above studies still included $10^7$~M$_\odot$ BHs for which the relation was found to be valid. Of course nonetheless the exact limit of validity will need to be further established both through theoretical and observational investigations.  

The implications of such high formation efficiencies ($\epsilon_{\rm BH} \sim 50\%$) first suggests that in the densest high-redshift environments, the central BH is not merely a byproduct of galaxy evolution but a dominant structural component, potentially explaining the anomalously high $M_{\rm BH}/M_{\rm gal}$ ratios observed in LRDs \citep{JUODZBALIS2024}. Second, this scenario alleviates the need for continuous super-Eddington accretion to grow the BH from stellar mass. Consequently, these massive BHs can exist in a quiescent or low-accretion state, naturally reconciling their large inferred masses with the strict X-ray non-detections reported in stacking analyses.

\subsection{X-ray background data and black hole emission}

We collected the X-ray data from the stacking procedure by \citetalias{SACCHI2025}, where they employed data from the Chandra Deep Field
South (CDF-S). The total exposure time, using the
data from all of the available sources, amounts to
$\approx 400$ Ms, leading to an unprecedented sensitivity of $\sim4 \times 10^{-18}$~${\rm erg\, s^{-1}\,cm^{-2}}$.

The upper limits of the detections are provided at the $3\sigma$ level, where $\sigma$ represents the standard deviation of the background noise. For the soft band ($0.3-2$\,keV) and the hard band ($2-7$\,keV) the upper limits are $<4.7\times 10^{-18}$ and $<1.3\times 10^{-17}$, respectively, in units of ${\rm erg\, s^{-1}\,cm^{-2}}$.

We model the intrinsic X-ray emission of the SMBHs formed in our model. We assume a power law with a high-energy exponential cutoff  \citep[e.g.,][]{YANG2020}
\begin{equation}
    f_{E} \propto E^{-\Gamma+1}\, \exp{(-E/E_{\rm cut})}, \label{eq:Flux}
\end{equation}
where $\Gamma$ is the photon index. The value of $\Gamma$ is usually set at $\approx 1.8-1.9$ \citep[e.g.,][]{PICONCELLI2005,YANG2016,LIU2017,YANG2020} and $E_{\rm cut}=300$\,keV, for Seyfert galaxies \citep[see][]{RICCI2017}. However, the weak emission of LRDs in the X-ray band is consistent with a steeper value of $\Gamma\approx 2.4\pm 0.1$ \citep[e.g.,][]{ZAPPACOSTA2023} or with the presence of a very low-energy cutoff $E_{\rm cut} \approx 20$\, keV. We adopt the canonical value $\Gamma=1.9$ with a low-energy cutoff of $E_{\rm cut}=20$\, keV. We also checked the possibility of $\Gamma=2.4$ finding no difference in our results.

The normalization constant of Eq.\,\ref{eq:Flux} is set such that the integrated X-ray luminosity corresponds to a fraction of the Eddington luminosity of the BHs, i.e.,
\begin{equation}
    L_X = \underbrace{ f_{\rm occupation}^{\rm BH}U_{\rm duty} \epsilon_{\rm Edd} \epsilon_X}_{k}  L_{\rm Edd},\label{eq:totalXrayLum}
\end{equation}
where $U_{\rm duty}$ is the duty cycle , $\epsilon_{\rm Edd}$ is ratio of bolometric luminosity to Eddington luminosity ($L_{\rm Edd}=1.26\times10^{38}\left(M_{\rm BH}/{\rm M}_\odot\right)~{\rm erg\,s^{-1}}$), $\epsilon_X$ the fraction of the bolometric luminosity  emitted in X-rays, and finally $f_{\rm occupation}^{\rm BH}$ is a factor that considers the possibility that not all the galaxies form a BH.

Our approach constrains the product $k=f^{\rm BH}_{\rm occupation}U_{\rm duty}\epsilon_{\rm Edd}\epsilon_{X}$ rather than individual parameters. It ranges over $0<k\leq 0.3$. Here, $0$ is the lower limit of no emission. The upper bound corresponds to a `Quasar Growth' scenario—necessary to grow massive BHs at $z>5$—implying, for example, a near-unity occupation fraction with an active duty cycle of $100\%$ at $\epsilon_{\rm Edd}=0.3$, or alternatively, short bursts of super-Eddington accretion ($\epsilon_{\rm Edd} \sim 1$) with a $30\%$ duty cycle. The conversion to observable X-ray flux is then modulated by $\epsilon_{\rm X}$, which we adopt from luminosity-dependent bolometric corrections \citep{DURAS2020} to account for the spectral softening at these high accretion rates.

As the parameter $k$ in our model represents the effective time-averaged accretion activity, physically, this degenerates the Eddington ratio ($\epsilon_{\rm Edd}$) and the accretion duty cycle ($U_{\rm duty}$), such that $k \propto \epsilon_{\rm Edd} \times U_{\rm duty}$.
For X-ray stacking analysis, which averages the flux over the population, these two factors are often indistinguishable; a source accreting continuously at $1\%$ Eddington luminosity produces a similar stacked signal to a source accreting at $100\%$ Eddington luminosity with a $1\%$ duty cycle. However, the duty cycle has important implications for mass growth. To grow from a small seed to $10^8 $\,M$_\odot$, a BH requires not just a high accretion rate, but also a high duty cycle ($U_{\rm duty} \sim 1$).

We include an absorption model to take into account the attenuation of the emission due to presence of column densities, which we describe via a suppression factor, i.e., $\exp(-\sigma_{\rm tot}(E)N_{\rm ave})$, where $\sigma_{\rm tot}(E)$ is the total cross section as a function of the energy of photons (see Eq. \ref{eq:totalCrossSection}) and $N_{\rm ave}$ is the average column density. The total cross section is given by 
\begin{equation}
    \sigma_{\rm tot} (E) = \sigma_{\rm H}(E)+\sigma_{\rm He}(E)+\left(\frac{Z}{Z_\odot}\right)\sum_{i>2} A_{\odot}^{i}\,\sigma_{i}(E),  \label{eq:totalCrossSection}
\end{equation}
with $Z$ the metallicity, $A^{i}_\odot$ are the Solar abundances taken from \cite{WILMS2000}. The cross sections for the metals ($\sigma_i(E)$) are obtained using the fits provided by \citet{VERNER1996}. Finally, the observed flux in a certain band is given by 
\begin{equation}
    F_X \propto {\int{\rm d}E_{\rm obs}\, f_{E}(E_{\rm emi}) \exp(-\sigma_{\rm tot}(E_{\rm emi})N_{\rm ave}) \over 4\pi D_L^2(z)(1+z)},
\end{equation}
where $D_L(z)$ is the cosmological luminosity distance, the factor $(1+z)$ corrects the flux for the redshifting of the photons, and the energy interval is stretched by cosmic expansion.

\section{Results} \label{sec:results}
Here, we describe the main results of our work. In Section \ref{sec:collisions} we show the results for the expected BH population based on the collision-based formation scenario. Subsequently, we provide the constraints on this population from the upper limits on the X-ray flux in Section \ref{subsec:XrayResults}. For all the calculations we adopt the conservative scenario in which BH masses cannot exceed $10^7$~M$_\odot$ as described in Section \ref{sec:BHFormation}.

\subsection{Expected black hole population from the collision-based  scenario}\label{sec:collisions}

\begin{figure}[!h]
    \centering
\includegraphics{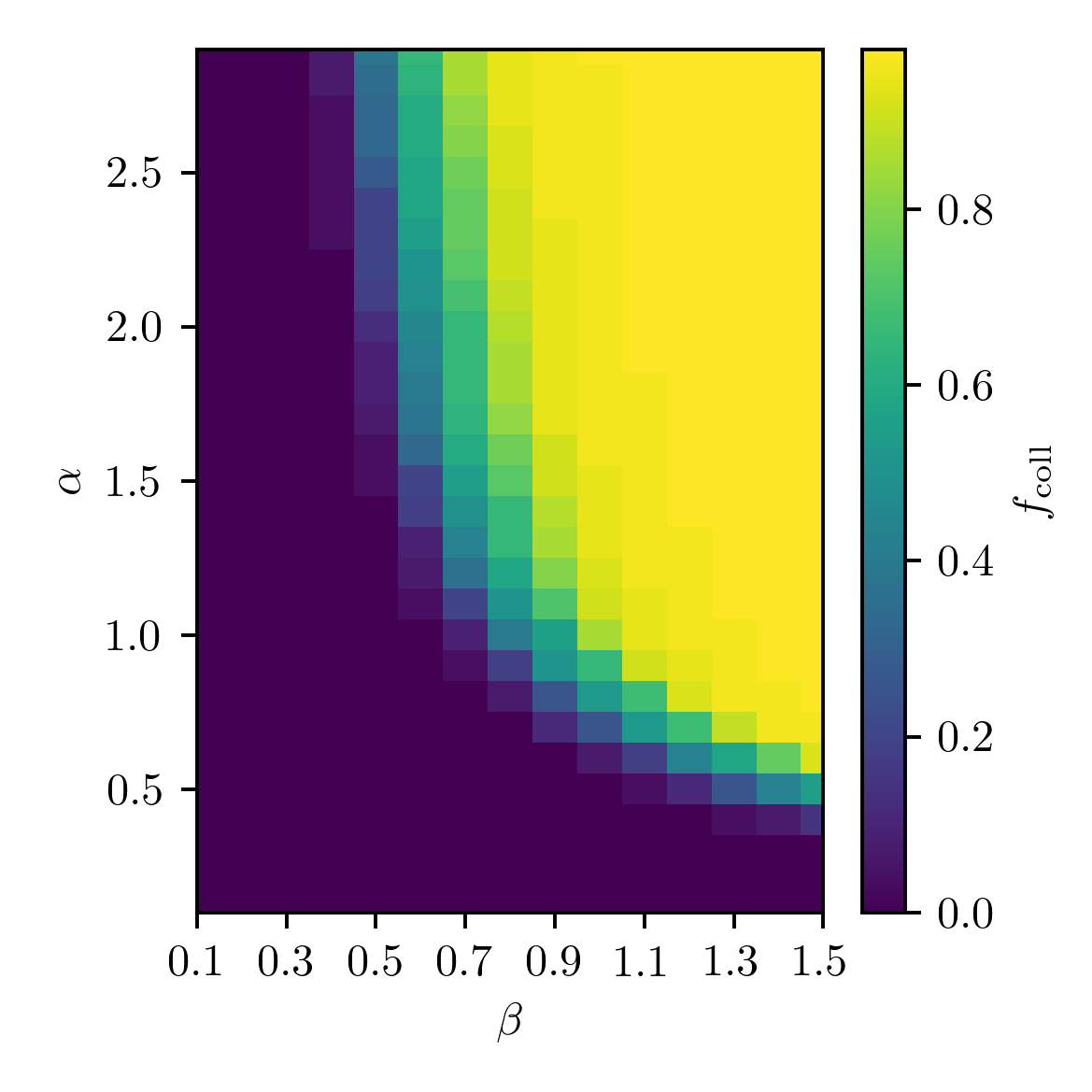}
    \caption{Fraction of galaxies where BH formation is driven by the condition $t=t_{\rm 
coll}$, as a function of the power-law exponents $\alpha$  and $\beta$ . The x-axis represents the exponent in the mass-radius relation $R_{\rm gal}\propto M_{\rm gal}^{\beta}$ and the y-axis represents the exponent in the mass-time relation $M_{\rm gal}\propto t^{\alpha}$.
The colorbar shows $f_{\rm coll}$ indicates the percentage of galaxies in which BH formation specifically satisfies $t=t_{\rm coll}$.} \label{fig:fractionBHFormed}
\end{figure}

We briefly recall here that, for galaxies where we reach the condition $t=t_{\rm coll}$ during their mass (size) evolution, we calculate the expected mass of the SMBH at that time; when it is not the case we calculate the product $\epsilon_{\rm BH}M_{\rm gal}$ at the time corresponding to the age of the stellar system, as defined via the observed redshift. Here, we first check for how many systems the condition $t=t_{\rm coll}$ has been fulfilled, asserting them with the parameters $\alpha$ and $\beta$, which are the power-law exponents that describe the time evolution of the mass of the LRDs and their adopted mass-radius relation. In Fig. \ref{fig:fractionBHFormed}, we provide a color-coded map of the fraction $f_{\rm coll}$ of systems that reach that condition as a function of the parameters $\alpha$ and $\beta$. We note that the results depend predominantly 
on the $\beta$-parameter. We obtain high fractions of such galaxies using $\beta\gtrsim0.7$, even if in some cases the condition is still reached for $\beta\sim0.5$ (adopting $\alpha\gtrsim1$). Here, $\alpha=1$ corresponds to a constant Star Formation Rate (SFR) while $\alpha\gtrsim1$ indicates increasing SFR with time.
The constraints on $\beta$ are somewhat more difficult to reconcile; in observed NSCs we obtain a typical relation with $\beta\sim0.5$. For early-type galaxies, typical values are $\beta=0.55-0.6$ \citep{SHEN2003}, while for late-type galaxies typical values are $\beta=0.22-0.27$ \citep{SHEN2003,LANGE2015}. If the mass-radius relation of LRDs is somewhat similar to spirals  or NSCs and the SFR increases with time, it is thus conceivable that the condition $t=t_{\rm coll}$ will be reached at least in some systems, allowing for particularly high efficiencies to form SMBHs. If this condition is still not reached, a BH may still form, but perhaps with an efficiency on the percent level or less $\sim1 \%$, given through Eq.~\ref{eq:blackHoleEfficiency}.

\begin{figure}[h!]
\centering
    \includegraphics{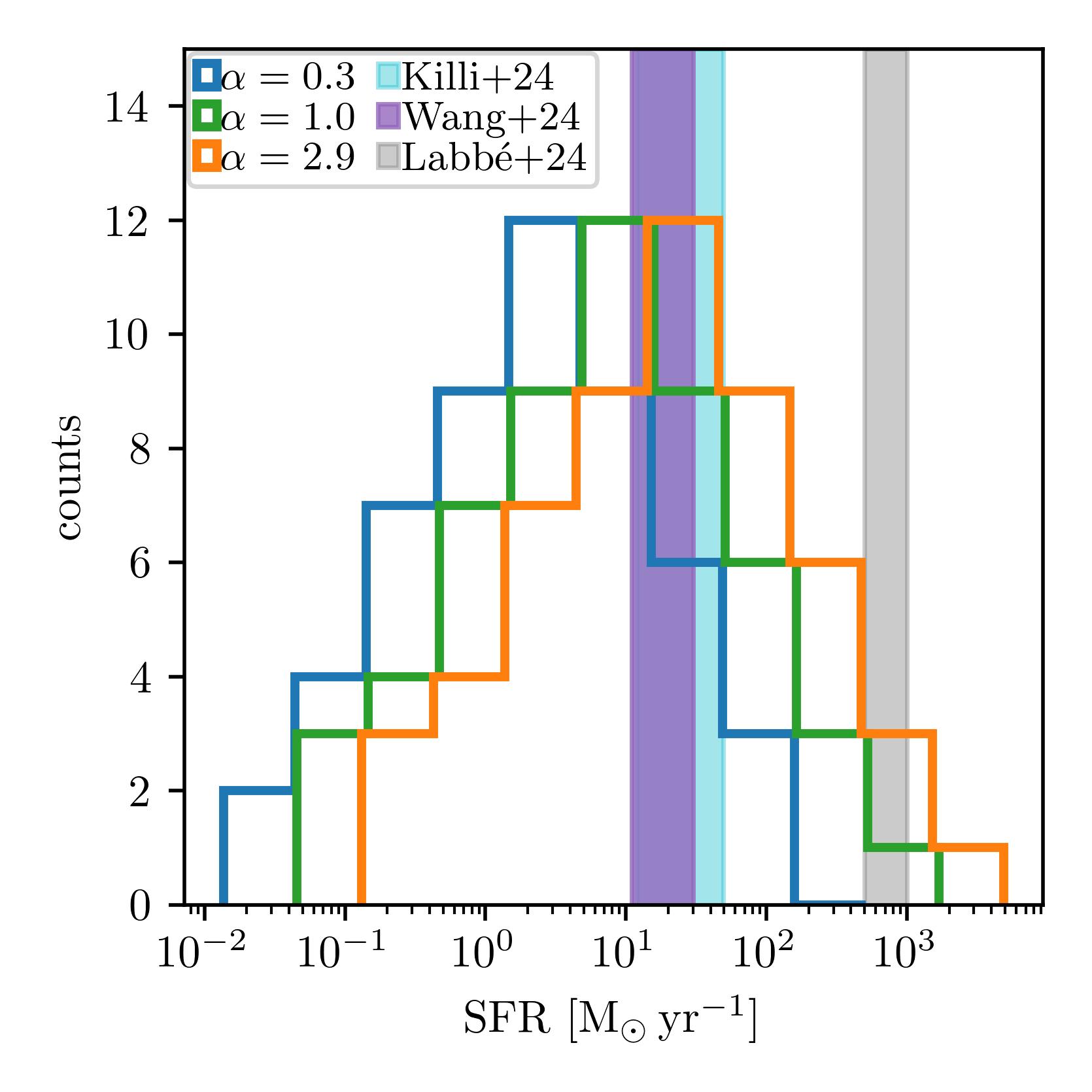}
    \caption{SFR distribution of our galaxy sample as function of $\alpha$. We show the distribution for $\alpha=0.3$ (blue), $1.0$ (green), and $2.9$ (orange). The cyan area shows the SFR range for the {J0647\textunderscore1045} galaxy at $z\sim4.53$ \citep{KILLI2024}, the violet area shows the range of three LRDs at $z\sim7-8$ \citep{WANG2024}, and in gray an extreme case of a LRD at $z\sim 4.47$  with a high SFR \citep{LABBE2024}.}
    \label{fig:SFRs}
\end{figure}

 In Fig. \ref{fig:SFRs} we show the SFR according to our model for the  galaxies in our sample for different values of $\alpha$. Specifically, we show the distribution for $\alpha=0.3, 1.0$ and $2.9$ (blue, green, and orange solid lines, respectively). For comparison, we include observational SFR estimations, in cyan we show the area corresponding to the SFR of a LRD at $z\sim 4.53$, the lower limit of the SFR is obtained from the [OII] line corrected by $A_V = 0.9 \pm 0.4$ with a mean value of $22\pm 13$~M$_\odot$\,yr$^{-1}$, the $H\alpha$ (narrow) line (corrected also by $A_{V}= 0.9 \pm 0.4$) provides a slightly higher value of $49^{+19}_{-13}$~M$_\odot$\,yr$^{-1}$. However,  the intrinsic luminosity at $L_{\rm 1500,UV}$ at $1500$ \r{A} from the UV continuum power-law fit suggest an SFR of $12\pm0.3$\, M$\odot$\,yr$^{-1}$ \citep{KILLI2024}.
In violet, it is shown the SFR of three LRDs. The averaged SFRs (taking into account the last $\sim 100$\,Myr) are $30.69^{+21.15}_{-13.37}$, $16.80^{+4.69}_{-4.36}$, and $11.06^{+7.24}_{-5.57}$~M$_\odot$\,yr$^{-1}$ \citep{WANG2024}. In gray, we show a LRD with the highest SFR estimated, the galaxy is observed at $z=4.47$ and its SED suggest that stars dominates the continuum, which self consistently explain the lack of X-ray detection and the lack of a hot dust upturn in the mid-infrared. The main issue arises when explaining the $H\alpha$ luminosity, a SFR of $500-1000$~M$_\odot$\,yr$^{-1}$ is required to produce sufficient ionizing photons for such emission \citep{LABBE2024}.  As expected, higher values of $\alpha$ (defined in Eq. \ref{eq:mass}) shift the distribution to the right. The peak of the distribution for $\alpha=1.0$ and $2.9$, are overall consistent with the estimated SFRs from the literature. Combining the information from Fig. \ref{fig:fractionBHFormed} and  Fig. \ref{fig:SFRs}, if we confirm from observations that LRDs are compact and massive stellar systems built-up in a short period of time, LRDs are more likely perfect places for the formation of massive BHs as result of stellar collisions.

\begin{figure}
\centering
\includegraphics[width=\hsize]{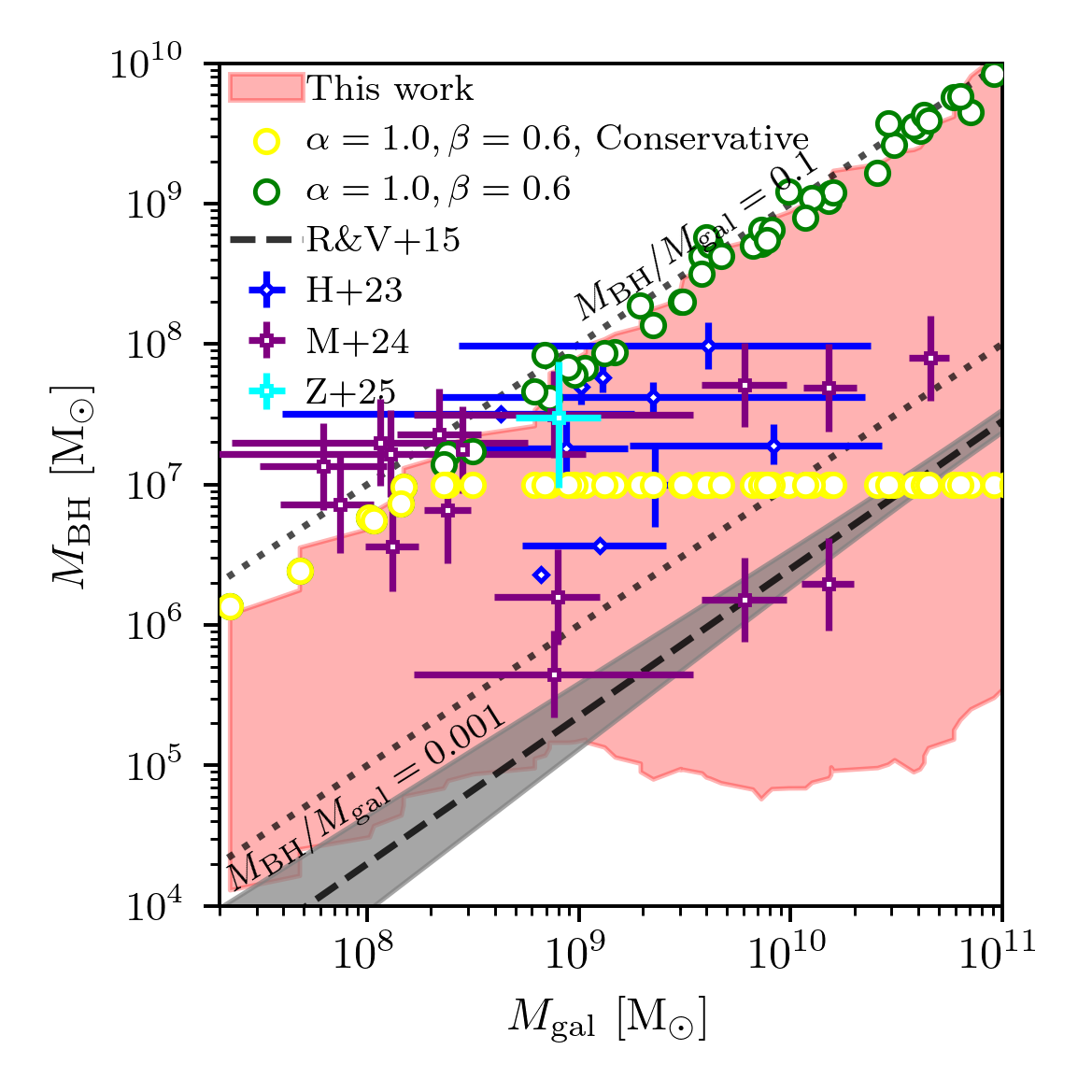}
\caption{The figure compares the $M_{\rm BH}-M_{\rm gal}$ relation from this work with observational  results. All masses are given in solar masses (M$_\odot$).  The red filled area represents the parameter space spanned by the  BH and host galaxy masses from this work, for different combinations of $\alpha$ and $\beta$. In yellow we show the BH (and host galaxy) mass for $\alpha=1.0$ and $\beta=0.6$ with a maximum mass seed of $10^7$\,M$_\odot$. On the other hand, in green we show the masses without limiting the mass of the seed. The black dashed line and gray shaded area show the empirical  $M_{\rm BH}-M_{\rm gal}$   relation for local (z=0) galaxies \citep{REINES2015}. We include data points from high-redshift (z=4-7) observational studies from  \cite{HARIKANE2023} (blue open diamonds), data points from \cite{MAIOLINO2024} (purple open squares), and from \cite{ZHANG2025} (cyan square).  Dotted black lines indicate constant BH to galaxy mass ratios, labeled as $M_{\rm BH}/M_{\rm gal} =0.1$ and $0.001$.}\label{fig:blackHoleMasses}
\end{figure}

We now show the expected relation between the BH mass and galaxies in the framework of our models, in Fig.~\ref{fig:blackHoleMasses}. There, we indicate the range of the possible solutions for $\alpha$ between $0.1$ and $3$ and $\beta$ between $0.1$ and $1.5$. We highlight the results of a favorable case with $\alpha=1$ and $\beta=0.6$, for which the efficiencies to form massive BHs are rather high and we tend to produce SMBHs well above the observed relation for local Universe galaxies. Our conservative scenario in which the maximum allowed BH mass is $10^{7}$~M$_\odot$ as explained in Section \ref{sec:BHFormation} (yellow points) exhibits masses that scales with $M_{\rm BH}/M_{\rm gal}\approx 0.1$ for galaxies with masses up to $10^{8}$\, M$_{\odot}$, while the local correlation measured by \cite{REINES2015} scales as $M_{\rm BH}/M_{\rm gal}  \lesssim 0.001$ (see gray region in Fig.  \ref{fig:blackHoleMasses}). Our optimistic scenario is shown in Fig. \ref{fig:blackHoleMasses} (green dots) where we find that all the BH seeds follow the $M_{\rm BH}/M_{\rm gal}\approx 0.1$ relation. 
It is not completely unexpected that the model for $\alpha=1.0$ and $\beta=0.6$ scales as $M_{\rm BH}/M_{\rm gal}\approx 0.1$. In this model, $M_{\rm gal}/M_{\rm crit}\sim 10^{-2}-10^{-1}$, implying BH efficiencies in the order of $\epsilon_{\rm BH}\sim 6\times 10^{-2}-10^{-1}$ which results in $M_{\rm BH}\sim 6\times 10^{-2}-10^{-1}M_{\rm gal}$. We use the conservative scenario for the subsequent analysis in this paper.
The masses obtained in this scenario are comparable with the high-redshift data points from \citet{HARIKANE2023}, \citet{MAIOLINO2024}, and \citet{ZHANG2025}. Nonetheless, as indicated in the figure, the overall possible parameter space is much larger, and depending on the assumed history of the galaxy, the parameter space also includes possible solutions with SMBHs at much lower masses. It is important to stress that the predictions provided here are only based on the seeding model itself and do not yet take into account possible growth via accretion of even super-Eddington accretion scenarios. 

Considering the points above, we have constrained $\alpha\gtrsim1$ from the comparison of the model histories with the typical SFRs \citep{KILLI2024, WANG2024, LABBE2024}. To also have efficient collisions, $\beta$ should be of the order $0.55$. In case of a too low value of $\beta$, there would not be collisions, while a very high value is perhaps unplausible as the largest observed mass-radius relations have power-law indices of $0.6$ \citep{SHEN2003}. When considering the X-ray constraints on our sources, we will therefore use the above as the reference parameters for the dynamical model.

In Appendix \ref{appendix:Accretion} we show how subsequent accretion further impacts these final BH masses. We demonstrate that for models in which seeds form early (e.g., $t_{\rm form} \approx 0.1 \, t_{\rm age}$ for high $\beta$), a long timescale is available for mass growth (see Fig. \ref{fig:fractionTime}). Given the already large masses of our seeds, this subsequent growth must be constrained.
In Fig. \ref{fig:Rates}, we explore the parameter space of accretion by considering a low radiative efficiency BH model and a high-efficiency (rapidly spinning) model \citep[e.g.,][]{SHAPIRO2005}. We find that for low radiative efficiency BHs, even moderate sub-Eddington rates ($L/L_{\rm Edd}\sim 0.1$) can lead to unphysically large final masses that systematically overshoot the observed relations. In contrast, for high radiative efficiency BHs, the mass growth is far more gradual and self-regulated. In this scenario, a moderate sub-Eddington rate \citep[$L/L_{\rm Edd}\sim 0.1$ as observed in Type 1 AGNs in the COSMOS field,][]{TRUMP2009} is physically plausible mechanism, producing a typical growth factor of $\sim 10$. A low sub-Eddington rate (i.e., $L/L_{\rm Edd}\lesssim 0.01$) is also entirely viable, representing the regime for low luminosity AGNs \citep[e.g.,][]{HO2008}. While at least for the typical population, the X-ray constraints seem to limit such scenarios, we also emphasize that JWST has found a few objects with BH to bulge mass ratios of order $40\%$ \citep{Juodbalis2024}, which potentially could be explained in a scenario of both massive seed formation and subsequent efficient accretion.

\subsection{X-ray background prediction} \label{subsec:XrayResults}
In this Section, we present the resulting constraints from the comparison of the predicted emission in different scenarios with the observational constraints in the soft and hard X-ray bands. In the following subsections, we will discuss the constraints that can be derived on the $\alpha$ and $\beta$ parameters characterizing the history of the LRDs and their mass-radius relation, the average column density ($N_{\rm ave}$), metallicity ($Z$) as well as the $k$ parameter defined in Eq. \ref{eq:totalXrayLum}. As it is a large parameter space, it is clear that there are degeneracies in the possible constraints. Nonetheless the upper limits on the fluxes rule out some possibilities and with increasing information about these sources, perhaps more constraints can be alleviated in the future.

\subsubsection{Constraints on star formation history and mass-radius relation }

Already in Section \ref{sec:collisions}, we have constrained the possible parameter space for $\alpha$ and $\beta$ considering the typical observed star formation rates in LRDs as well as model-constraints from the requirement of having collisions. Here, we check if these are also compatible with the constraints from the X-ray background. For this purpose, it is important to carefully consider which parameters we can assume for column density, metallicity and activity level.

Observations by JWST in principle have detected broad emission lines in the rest-frame UV/optical, suggesting the broad-line regions to be relatively unobscured \citep{GREENE2024}. However, \citet{MAIOLINO2025} have argued that dense, dust-free clouds that are causing the broad-line emission could simultaneously provide shielding in the X-rays. As a result, we either have the case of very strong shielding (for which we are essentially unable to constrain the other parameters), or we are in the regime of low column densities. We focus here on the second regime and adopt a reference value of $N_{\rm ave}=10^{21}$~cm$^{-2}$ (we checked that the following results also hold in case of lower column densities). Measurements of the metallicity are not available so far, but we adopt here a reference value of $0.01$~Z$_\odot$ as a typical value for high-redshift galaxies  \citep[e.g.,][]{CURTI2024,MEYER2024, SANDERS2024}. In the regime of reasonably high metallicities, the product between column density and metallicity has the main effect with respect to the shielding due to the exponential dependence. The dependence on the activity parameter $k$ on the other hand only is linear. Therefore, the activity parameter will be harder to constrain, but we assume here a  typical value of $k=2.5\times10^{-3}$ for definiteness. We will explore variations in all these parameters in the following sub-sections.
    
\begin{figure}[h!]
    \centering
    \includegraphics[height=0.85\textwidth,
  keepaspectratio]{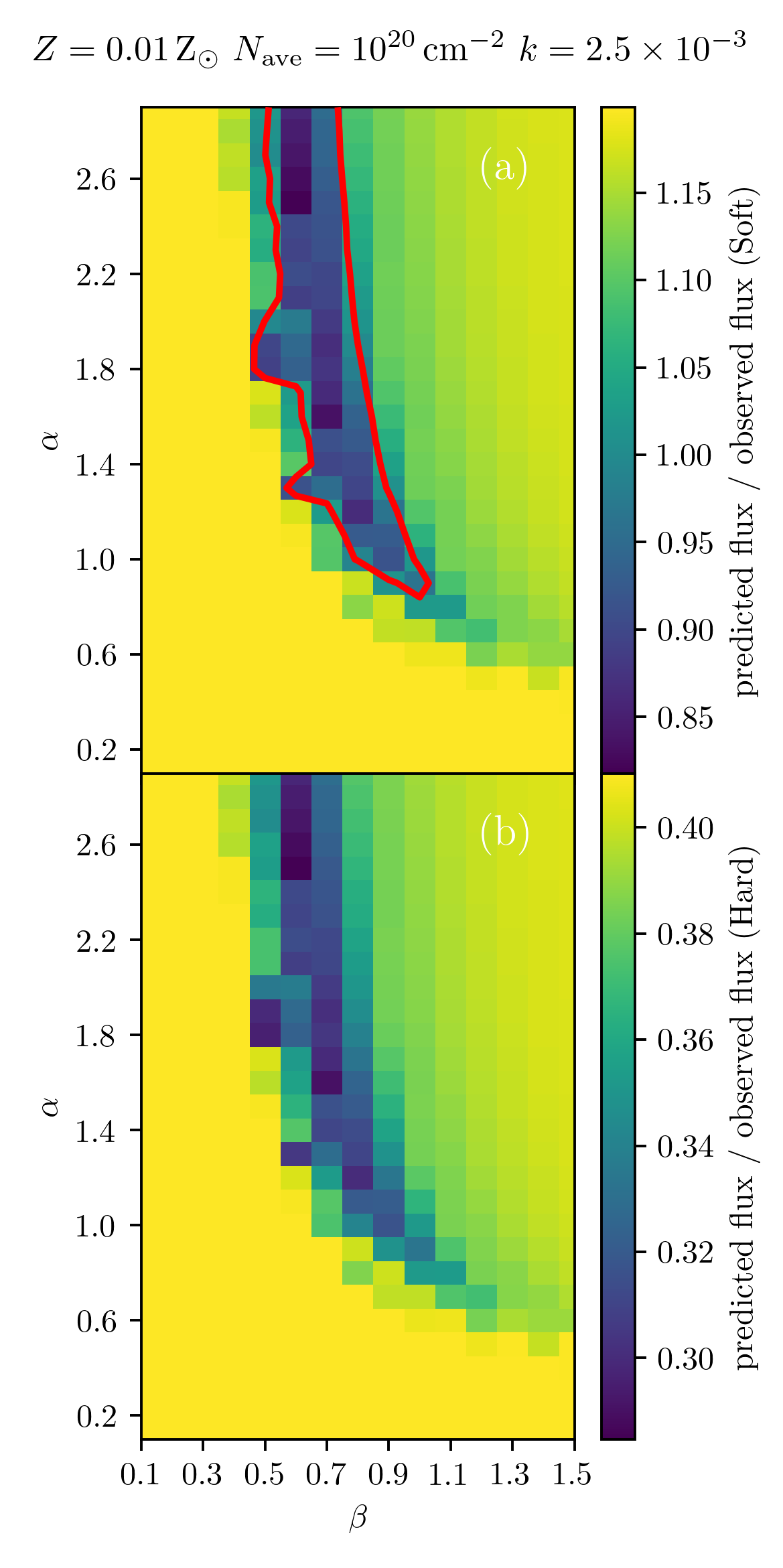}
    \caption{
    Panel (a) shows the soft X-ray band ($0.3-2$ keV) and panel (b) shows the hard X-ray band ($2-7$ keV). The colorbar shows the ratio between the predicted and the observed flux for the soft and hard bands.
    The allowed parameter space for $\alpha$ and $\beta$ is the region where the ratio of the predicted flux to the observed flux is less than 1, which satisfies the soft X-ray and hard X-ray observational constraints, respectively. Red line represents ratio between predicted and observed flux equals to one.}\label{fig:alphaBetaConstrains}
\end{figure}

 The results for the soft band and hard band are provided in Fig.~\ref{fig:alphaBetaConstrains} (panel a and panel b, respectively), where the color scale indicates the ratio of predicted versus observed flux as a function of $\alpha$ and $\beta$. The allowed parameter space is for which the ratio  of predicted versus observed flux is less than one (red line shows where the ratio is equals to one), corresponding to the allowed region in the parameter space for the given choice of column, metallicity and $k$-parameter. The X-ray constraints indicate the allowed parameter space for $\alpha$ and $\beta$, implying again $\alpha\gtrsim1$ and $\beta<0.6$. The X-ray constraints are thus compatible with the requirements of the model. In the following subsections, we will employ the constraints on $\alpha$ and $\beta$ derived here, of course still under the assumption that obscuration is non relevant (as clearly no constraints could be obtained in a heavily obscured regime).
 
 The qualitative behavior of the hard band is very similar to the soft band. This is a generic result that the constraints obtained from the soft band are more tight, and we will therefore mostly focus on these similarly to \cite{YUE2024} showed that the soft band is more sensitive to variations, e.g., in column density, than the hard band.

\subsubsection{Metallicity and average column density} \label{subsubsec:ZandN}

The amount of shielding of galaxies is an important parameter as the absorption of the X-ray flux can significantly alleviate possible constraints. It depends both on the average column density in the sources but also on the metallicity, as the cross section for X-rays can be significantly increased in the presence of heavy elements. 

We consider again our reference model for the collision-based BH formation scenario with $\alpha=1$ and $\beta=0.6$ (considering that collisions must be sufficiently effective while maintaining still a realistic value of $\beta$) to assess the predicted flux and how it compares to the upper bounds.  

\begin{figure}[h!]
    \centering
\includegraphics[height=0.75\textwidth,
  keepaspectratio]{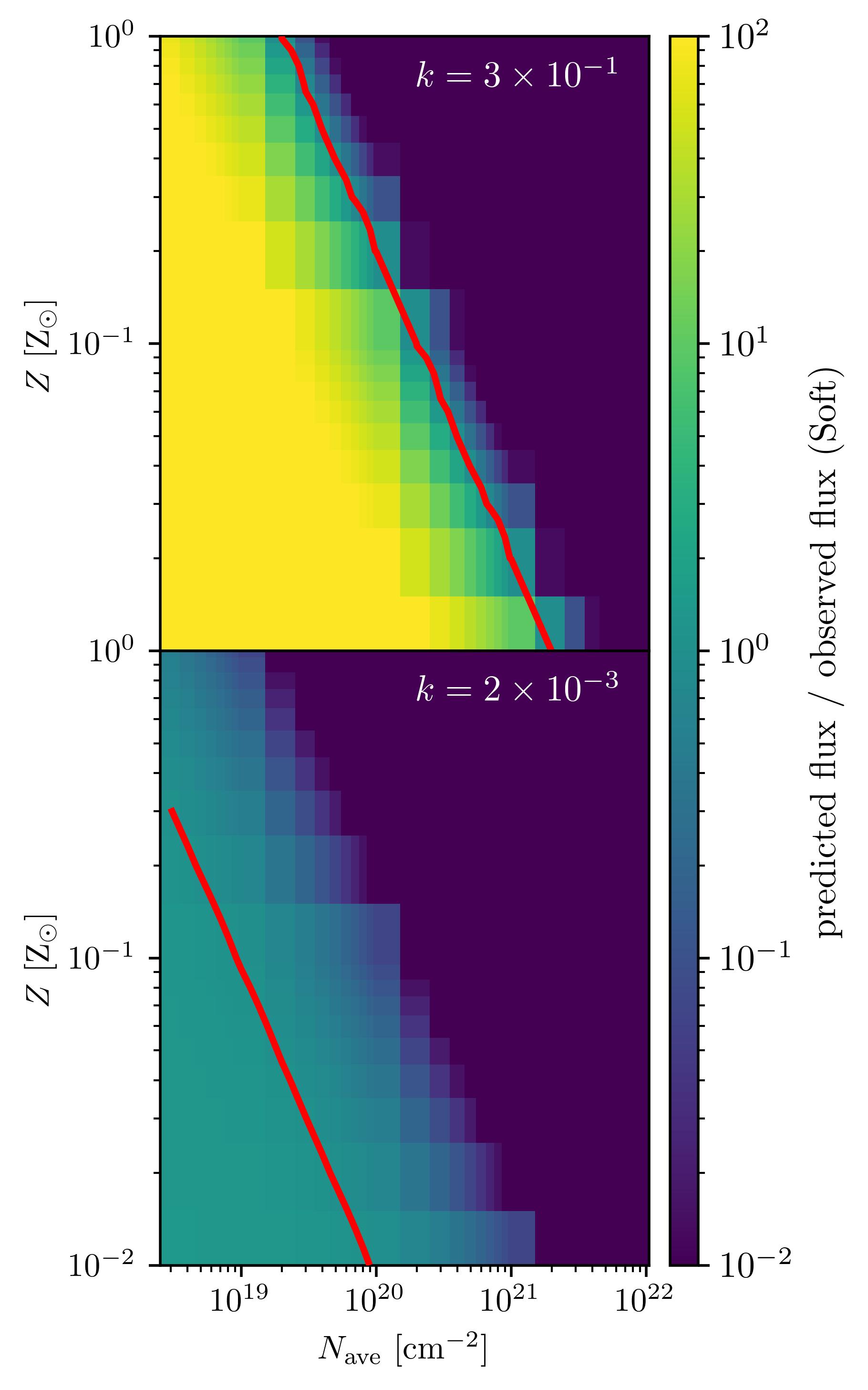}
    \caption{The colorbar shows the ratio between the predicted flux from our BH population and the observed limits in the soft band (0.3-2 keV). The  solid red line denotes the curve where the ratio between the predicted flux and the observed one is equal to one. The y-axis shows the metallicities from solar to $Z=0.01~{\rm Z}_\odot$. The x-axis is the average column density $N_{\rm ave}$ in units of ${\rm cm^{-2}}$. We adopted $k=0.3$ as an example of an extreme case of high activity (top panel) and a more conservative value $k=2\times10^{-3}$ (bottom panel).}
    \label{fig:ZvsNave}
\end{figure}

In Fig. \ref{fig:ZvsNave}, we  present the results for a value of $k=0.3$, corresponding to a case with high Eddington ratio, duty cycle and BH population fraction in the top panel, and we show the results of a value equals to $k=2\times10^{-3}$ for a more conservative scenario (bottom panel). The colorbar indicates the ratio of predicted versus observed flux in the soft band. So far, no metallicities of LRDs have been inferred from observations. Furthermore, observations at low and high-redshift suggest that the range of metallicities is not constrained by redshift. Indeed, at high-redshift, highly evolved objects like AGNs also show close-to-solar metallicities \citep[e.g.,][]{NAGAO2006}. 

For the high activity model (top panel), the constraints correspond to a triangular shape, where at low metallicities column densities of $\sim2\times10^{21}$~cm$^{-2}$ will be required, while with increasing metallicity even lower columns of $\sim2\times10^{19}$~cm$^{-2}$ can be sufficient. We have checked that the corresponding results for the hard band are very similar. When we explore a lower value of $k$, the constraints become tighter and the required column density for shielding decreases compared to the previous result. A column density of $\sim9\times10^{19}$~cm$^{-2}$ is required to shield the emission at $Z=10^{-2}$\,Z$_\odot$. On the other hand,  at solar metallicities the required density is in the order of $\sim3\times10^{18}$~cm$^{-2}$. 

We note that in the high-activity regime $(k=0.3)$, we recover tighter constraints than in the conservative scenario ($k=2\times 10^{-3}$).  This may appear counter-intuitive, as high-Eddington sources are known to exhibit steeper X-ray spectra and larger bolometric corrections (i.e., lower intrinsic X-ray efficiencies, $\epsilon_X$), making them relatively 'X-ray weak' compared to their bolometric output \citep[e.g.,][]{DURAS2020, MADAU2024}. However, our results demonstrate that the massive increase in the total bolometric budget ($\epsilon_{\rm Edd}$) dominates over the suppression of X-ray efficiency. While $\epsilon_X$ decreases by a factor of $\sim 3-4$ as the accretion rate rises from $\epsilon_{\rm Edd} \sim 10^{-3}$ to $\epsilon_{\rm Edd} \sim 0.3$, the bolometric luminosity increases by two orders of magnitude. Consequently, the net absolute X-ray luminosity ($L_X \propto \epsilon_X \cdot \epsilon_{\rm Edd}$) remains significantly higher in the high-$k$ scenario. Because these sources are intrinsically brighter in X-rays despite their steeper spectra, they are more easily ruled out by the stacking upper limits, resulting in the tighter constraints observed in Fig. \ref{fig:ZvsNave}.

\subsubsection{Black hole activity and average column density}\label{subsubsec:kVSNave}

\begin{figure}[h!]
\centering
\includegraphics[height=0.75\textwidth,
  keepaspectratio]{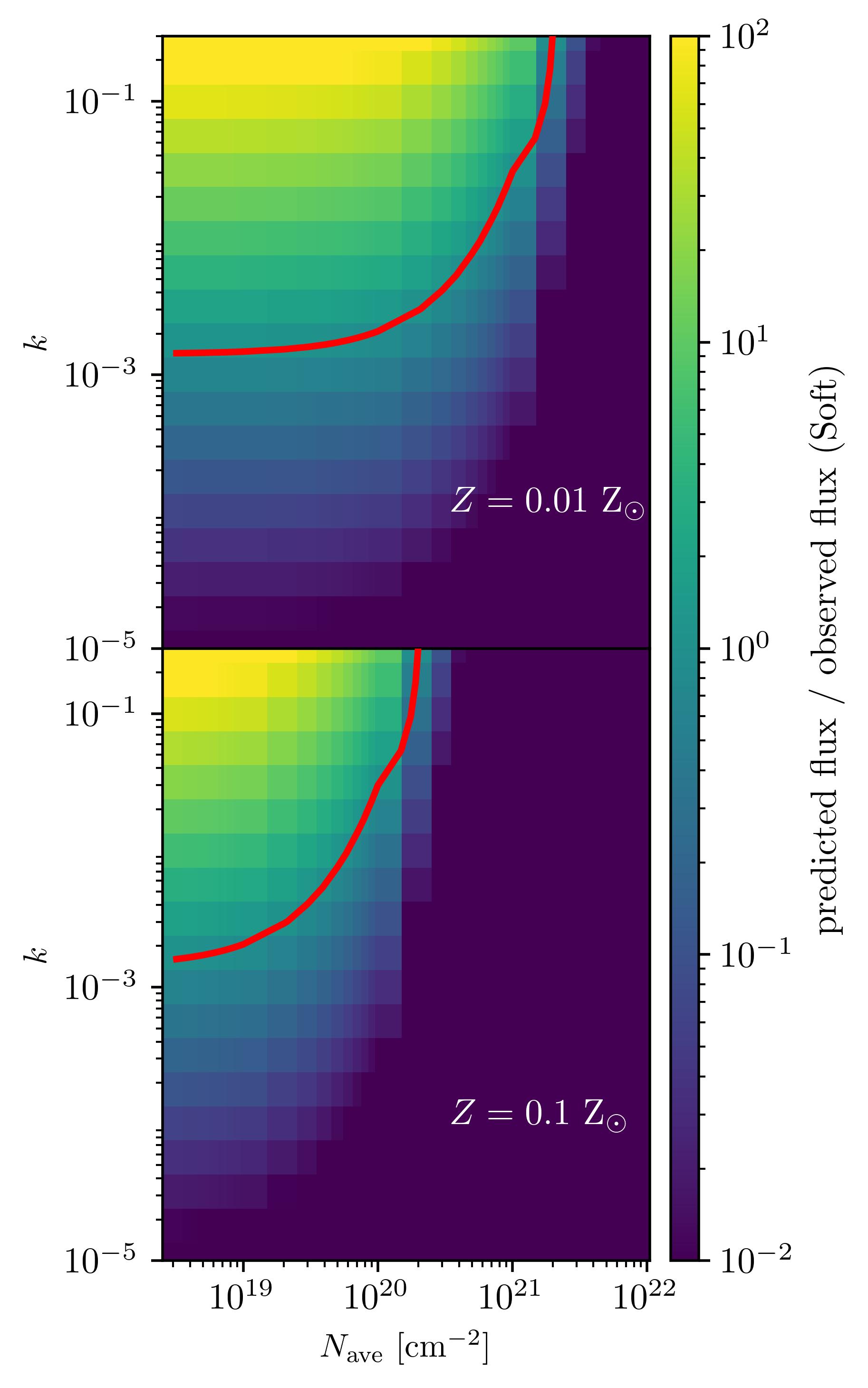}
    \caption{The colorbar shows the ratio between the predicted flux from our BH population and the observed limits in the soft band (0.3-2 keV). The y-axis shows the possible values for $k=f_{\rm occupation}^{\rm BH}U_{\rm duty}\epsilon_{\rm Edd}\epsilon_{ X}$ assuming a metallicity equals to $Z=0.01~{\rm Z}_\odot$ (top panel) and $Z=0.1~{\rm Z}_\odot$ (bottom panel) . The x-axis is the average column density $N_{\rm ave}$ in units of ${\rm cm^{-2}}$. The solid red line denotes the curve where the transition from over predicted takes place.}
    \label{fig:kVSN}
\end{figure}

In this subsection, we assess the joint constraints on the gas column density and the activity of the BH population, which we summarized in the $k$-parameter defined as $k=f^{\rm BH}_{\rm occupation}U_{\rm duty}\epsilon_{\rm Edd}\epsilon_X$, and the average column density for shielding.  We adopt our reference model with $\alpha=1$ and $\beta=0.6$. The stronger constraints are again obtained in the soft band, which we provide in Fig. \ref{fig:kVSN}. We find that for a metallicity of $Z=0.1$~Z$_\odot$, average columns densities of the order of $2\times 10^{20}$~cm$^{-2}$ (or larger) definitely obscure the X-ray emission from the BH population independent of the value of $k$. A very similar conclusion can be obtained from the analysis of the hard band. For comparison, we also provide the constraints for a lower metallicity case with  ($Z=0.01~{\rm Z}_\odot$). Here, larger averaged columns are required to obscure the X-ray emission. For $k\gtrsim  10^{-3}$, our BH population is compatible with column densities of the order  $\sim 2\times10^{21}$~cm$^{-2}$.

\subsubsection{Black hole activity and metallicity}\label{subsubsec:kVSZ}

To assess the joint constraints on BH activity (parametrized through $k$) and the metallicity of the shielding column, we again adopt our reference model, with a generic column of $N_{\rm ave}=10^{20}$~cm$^{-2}$. 

We start the analysis with an unobscured region, i.e. as shown in Fig. \ref{fig:ZvsNave}, a region with $N_{\rm ave}=10^{20}$~cm$^{-2}$, where the resulting X-ray flux is exceeding the upper limits for the range of metallicities considered here, unless for values of $k\lesssim10^{-3}$.

In the case of the soft band,  the region restricted to $k>6\times 10^{-3}$  and $Z \lesssim 0.2\,{ \rm Z}_\odot$ overpredicts the upper limit of the flux by a factor of $\sim 100$. A low  average column density of $\sim 10^{20}$~cm$^{-2}$  thus also would then require $k\lesssim 6\times 10^{-3}$ for metallicities ranging from $1-0.01$~Z$_\odot$.

\begin{figure}[h!]
    \centering
\includegraphics[height=0.75\textwidth,
  keepaspectratio]{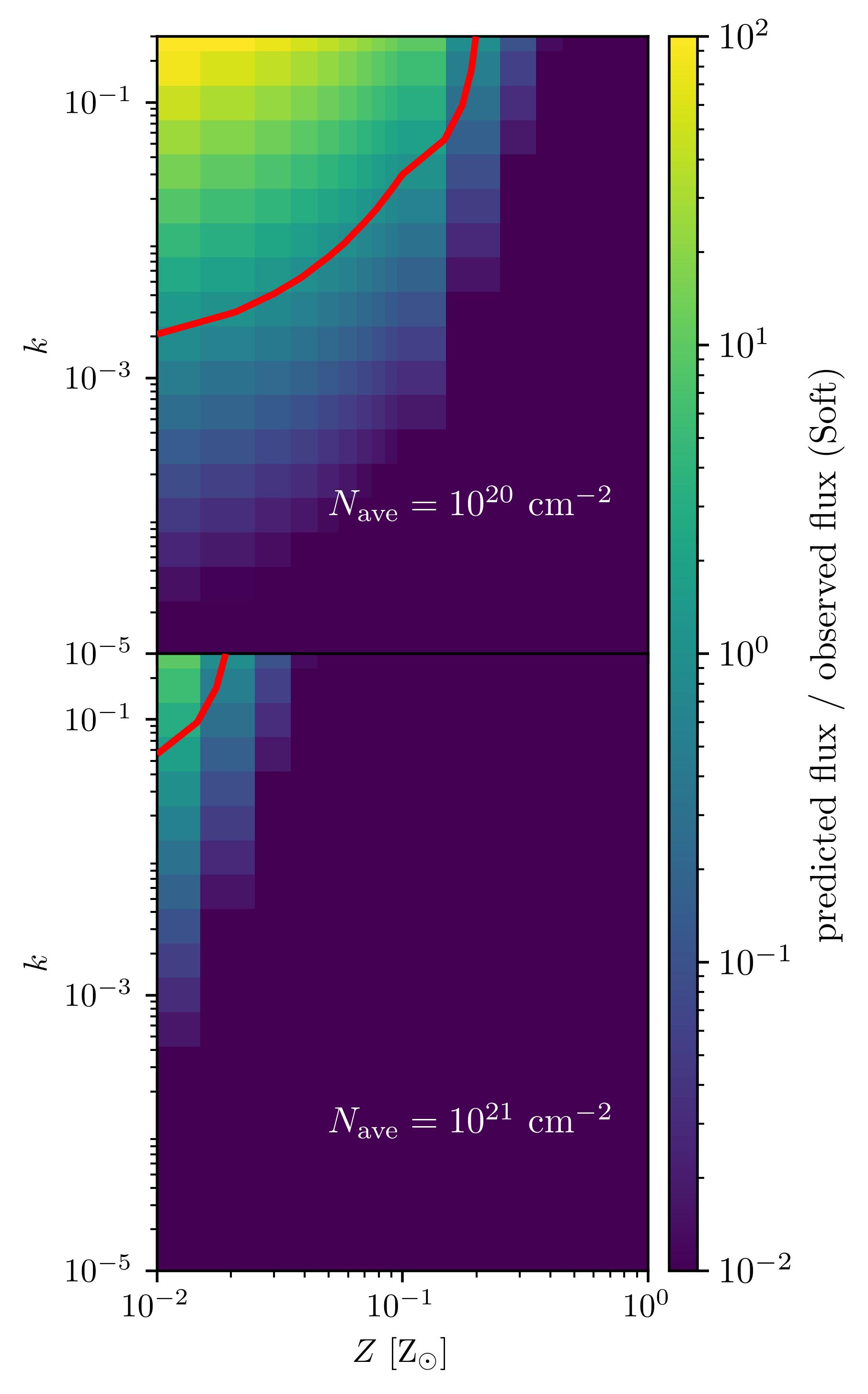}
    \caption{The colorbar shows the ratio between the predicted flux from our BH population and the observed limits in the soft band (0.3-2 keV). The  solid red line denotes the curve where the ratio between the predicted flux and the observed one is equal to one. The x-axis shows the metallicities from solar to $Z=0.01~{\rm Z}_\odot$. The y-axis is the $k$ parameter. We adopted average column density $N_{\rm ave}=10^{20}$\,${\rm cm^{-2}}$  (top panel) and  $N_{\rm ave}=10^{21}$\,${\rm cm^{-2}}$ (bottom panel).}
    \label{fig:ZvsNaveDiff}
\end{figure}

On the other hand, for high-redshift BHs we may expect larger values of the duty cycle and the Eddington ratio, even if the precise value of $k$ is unconstrained. The situation improves quite relevantly with a larger column density of $N_{\rm ave}=10^{21}$~cm$^{-2}$, for which even values of $k\sim0.3$ become compatible with the constraints in case that the metallicity is high. Even at lower metallicities this effect can likely be mitigated by larger gas columns increasing the amount of shielding. The overall constraints for this case are summarized in Fig.~\ref{fig:ZvsNaveDiff} as a function of metallicity and as a function of $k$.

\subsection{Comparison with IR inferred black hole masses}
\label{sec:IR_masses}
In this Section, we compare our predictions with BH masses inferred from the luminosities in the F444W band. We convert the apparent magnitudes to flux densities ($f_{\rm F444W}$) and subsequently to bolometric luminosities ($L_{\rm bol}$) using a Bolometric Correction (BC), such that $L_{\rm bol} = \text{BC} \cdot 4\pi D_L^2 \nu f_{\rm F444W}$.
Assuming the source radiates at a specific Eddington fraction, $\epsilon_{\rm Edd}$, we estimate the BH mass as:
\begin{equation}
    M_{\rm BH} = \frac{L_{\rm bol}}{\epsilon_{\rm Edd} \cdot (1.26 \times 10^{38} \text{ erg/s}/M_{\odot})}
    \label{eq:mbh_from_Lbol}
\end{equation}

We adopt three different BCs from \cite{GREENE2025}: $\rm BC=8$ (standard), ${\rm BC}=3.1$ (minimum), and ${\rm BC}=21.0$ (maximum), reflecting the uncertainty in the SEDs of LRDs. Exploring this range allows us to bracket the possible BH masses despite the lack of calibrated corrections for this population.

\begin{figure}[!ht]
    \centering
    \includegraphics[scale=0.95]{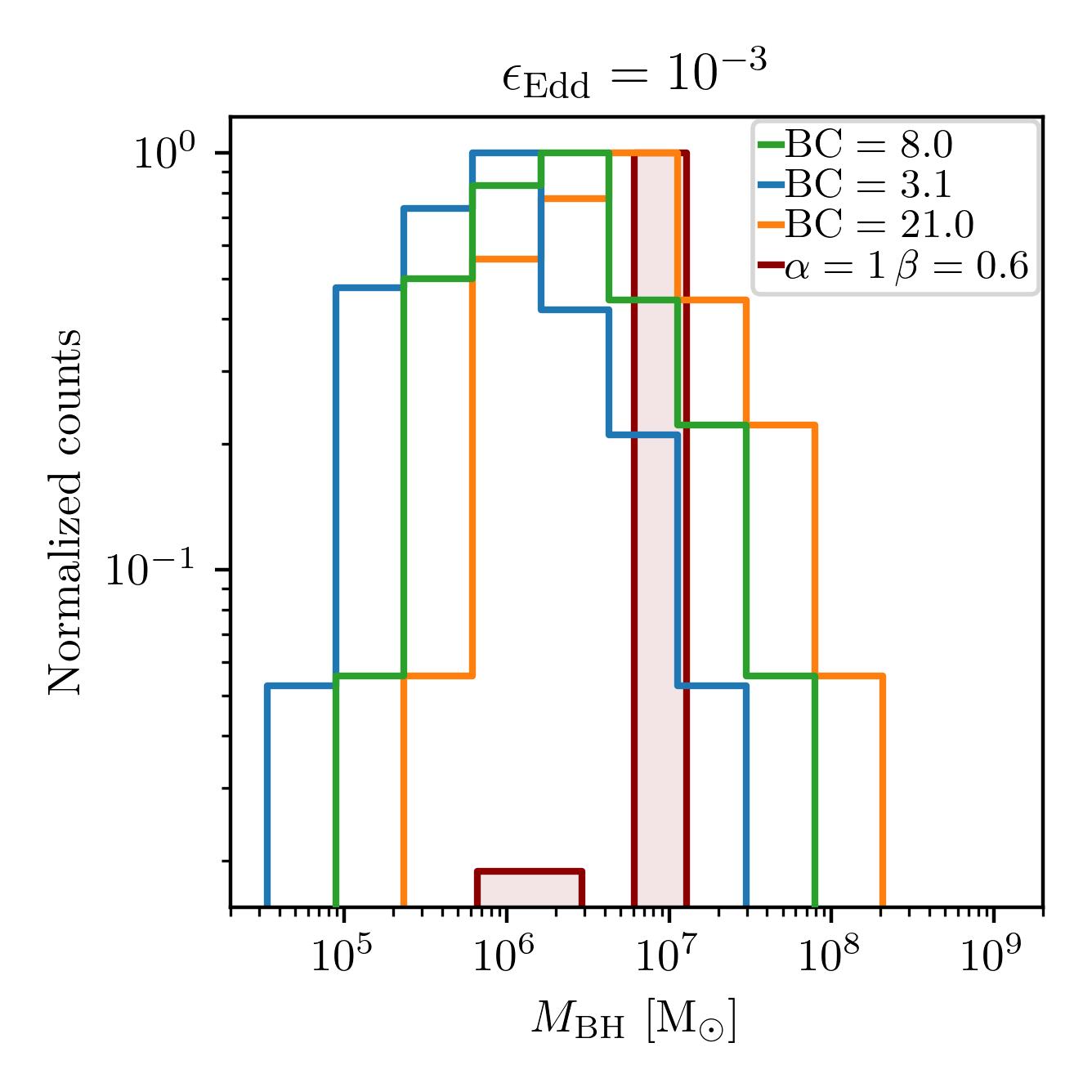}
    \\[-2.70ex]
    \includegraphics[scale=0.95]{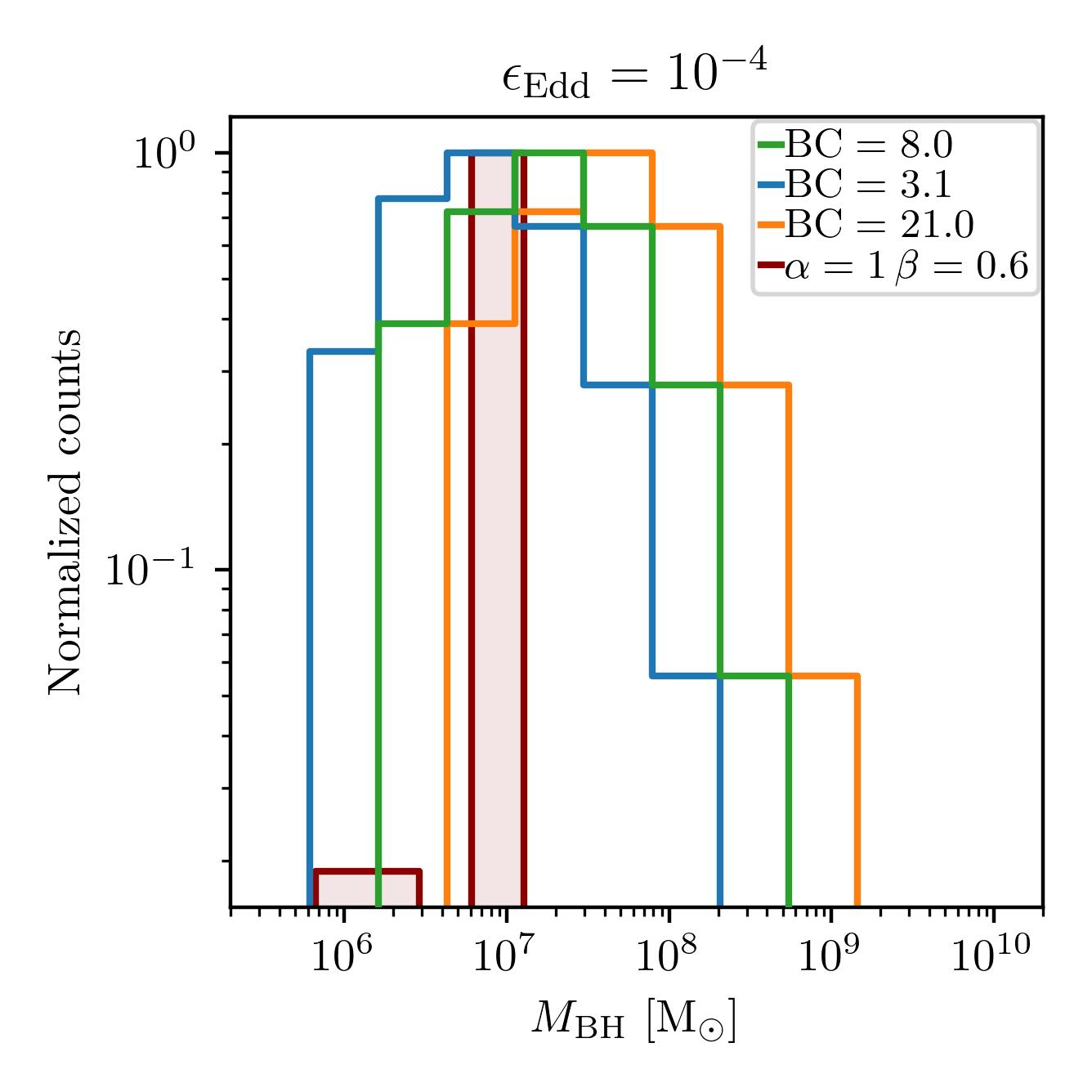}
    \caption{Distribution of BH masses inferred with the luminosities obtained from the $m_{F444W}$ magnitude, compared to our reference model $\alpha=1$ and $\beta=0.6$ (filled red histogram). Top panel assumes an Eddington fraction $\epsilon_{\rm Edd}=10^{-3}$. Bottom panel assumes $\epsilon_{\rm Edd } =10^{-4}$. The colored lines show the mass distributions for the standard bolometric correction (BC=8.0, green), the minimum (BC=3.1, blue), and the maximum (BC=21.0, orange).}
    \label{fig:bhMasses}
\end{figure} 

In Fig. \ref{fig:bhMasses}, we compare these IR-inferred masses with our reference collision model ($\alpha=1, \beta=0.6$, red histogram).
The top panel assumes a moderate Eddington fraction of $\epsilon_{\rm Edd}=10^{-3}$. In this scenario, the peak of our model distribution ($\sim 10^7\,{\rm M}_\odot$) is consistent with the bulk of the IR-inferred masses (green and orange lines). However, we explicitly note a discrepancy in the distribution widths: while the IR-derived masses span a broad range extending to $>10^8\,{\rm M}_\odot$, our model is strictly capped at $\sim 10^7\,{\rm M}_\odot$ due to the validation limit of the collisional channel.

The bottom panel ($\epsilon_{\rm Edd}=10^{-4}$) shows consistency with the minimum BC (blue line). We further explore the high-accretion regime ($\epsilon_{\rm Edd}=0.3$) in Appendix \ref{appendix:D}, finding that such high ratios would result in predicted BH masses significantly exceeding those estimated from IR observations. This reinforces our conclusion that the LRD population is best described by heavy seeds accreting at moderate-to-low efficiency or with low duty cycles.

\section{Summary and discussions}\label{discussion}

In this work, we provide estimates for the possible masses of SMBHs formed in LRDs via collisions. For this purpose, we present a toy-model assuming a power-law evolution of the stellar mass with time with a power-law index $\alpha$, as well as a mass-radius relation where the radius scales as mass to the power $\beta$. We included a BH formation scenario based on runaway stellar collisions using stellar systems with collisional timescales shorter than their age, which is motivated by previous studies of \citet{ESCALA2021,ESCALA2025}; \citetalias{VERGARA2023, VERGARA2024, VERGARA2025a, VERGARA2025b}. We find that the possibility to form massive BHs is particularly sensitive to the parameter $\beta$ ($R_{\rm gal}\propto M_{\rm gal}^{\beta}$), which should be of order $0.6$ or larger for the collision-based channel to be highly efficient. In this case, it is capable to produce SMBHs considerably above the BH mass - bulge mass relation observed in the local Universe \citep[e.g.,][]{REINES2015}, while consistent with high-redshift data such as \citet{HARIKANE2023, MAIOLINO2024}. Even for values of $\beta\sim0.6$, which is more typical for spiral galaxies in the local Universe \citep{SHEN2003}, high efficiencies to form massive objects can still be obtained, in particular when the value of $\alpha$ is above $1$, implying the SFR to increase as a function of time. The range of $\alpha$ and $\beta$  values has been determined from purely theoretical considerations requiring collisions to be efficient while keeping $\beta$ within a realistic range as well as from considering the typical observed star formation rates in LRDs. In addition, we checked that they are also compatible with the X-ray constraints in the regime of low column densities (in the case of a strongly obscured system as proposed by \citet{MAIOLINO2025}, no constraints can be obtained). We generally emphasize that the derived constraints from the X-ray background are only valid in the regime where obscuration is not significant, as otherwise clearly no relevant constraints could be obtained.

High-z galaxies are observed to have an increasing star formation history \citep{TOPPING2022}. Cosmological simulations using high-resolution radiation-hydrodynamic suggest that the SFR grows  exponentially \citep{PALLOTTINI2017,PALLOTTINI2023}. We note that even in cases where the SFR is constant ($\alpha=1$) or decreasing with time ($\alpha<1$), massive BHs are still likely to form though at a reduced mass, but of course their further growth would  still be possible via accretion, possibly even at super-Eddington rates \citep[e.g.,][]{INAYOSHI2016, TAKEO2019, REGAN2019}. There are observational constraints which seem to unfavorably affect super-Eddington accretion at the current stage of their evolution, though it may be too early to conclude about possible previous phases, where the main uncertainty is how long super-Eddington accretion lasts \citep{VOLONTERI2021}.

The intrinsic X-ray emission model includes an external attenuation from column densities ($N_{\rm ave}$) with different metallicities ($Z$).  We have introduced a parameter for the overall X-ray activity $k=f_{\rm occupation}^{\rm BH}U_{\rm duty} \epsilon_{\rm Edd} \epsilon_X L_{\rm Edd}$, which depends on the fraction of sources with a massive BH $f_{\rm occupation}^{\rm BH}$, the duty cycle $U_{\rm duty}$, the Eddington ratio $\epsilon_{\rm Edd}$, and the fraction of the bolometric luminosity emitted in X-rays $\epsilon_X$. We use this framework to derive the expected soft and hard X-ray emission from the sample of \citetalias{SACCHI2025}. We note that, although the radii of the LRDs in their sample have not been released to the public, we assume that they follow the same distribution as the sample of \citetalias{AKINS2024}. We then use a Monte-Carlo approach to assign initial radii to sources. This allows us to compare the expected emission to the upper limits that \citetalias{SACCHI2025} derived via stacking techniques from the Chandra Deep Field South (CDF-S), corresponding to a total exposure time of $400$~Ms. High X-ray activity levels, corresponding to large values of $k$, are feasible but require a larger column densities and/or higher levels of metal enrichment in these columns, as the heavy elements significantly increase the cross section for X-ray absorption.

Our collision-based BH formation scenario provides a natural explanation for the detection of over massive BHs and reconciles their presence with the reported weakness of X-ray signals without requiring high levels of obscuration. Indeed, for our most extreme case, where all the BHs are continuously accreting and emitting $30\%$ of their Eddington luminosity in X-rays, at solar metallicities  a column density of only $N_{\rm ave}\sim 3 \times 10^{18}$\,cm$^{-2}$ is required to be compatible with the observational constraints, while at lower metallicities ($Z=10^{-2}$\,Z$_\odot$) a column density around $N_{\rm ave}\sim 2 \times 10^{21}$\,cm$^{-2}$ is enough to reduce the X-ray emission. However, as explored in Appendix \ref{appendix:Accretion}, the subsequent growth should then be well-regulated to be compatible with the X-ray constraints, at least for the typical population. Indeed, a moderate sub-Eddington accretion rate ($L/L_{\rm Edd} \approx 0.1$) —consistent with typical Type 1 AGNs—provides a plausible explanation to consider the existence of accretion. Nonetheless, we also point out that JWST has even found some more peculiar sources with BH to bulge mass ratios of order $40\%$ \citep{Juodbalis2024}. In such cases, the combination of high seed masses plus very efficient accretion could be a promising way to explain the extreme outcomes in the more extreme sources. More extreme scenarios in which $\beta$ takes on higher values ($\beta \gtrsim 1.2$), resulting in significantly earlier seed formation times and consequently altering the requirements for the subsequent accretion history and duty cycles, are discussed in detail in Appendix \ref{appendix:Accretion}.

\cite{LATIF2025} recently explored the expected radio signatures of LRDs. Their analysis indicates that the radio flux from the AGN component generally dominates over the stellar component, often by factors of $10-100$, particularly for lower SFRs ($< 10\,{\rm M}_{\odot}$\,yr$^{-1}$). However, at higher SFRs ($\ge 10-30 \,{\rm M}_{\odot}$\,yr$^{-1}$), which our models and cited observations suggest are plausible for LRDs (Fig.~\ref{fig:SFRs}), the radio emission linked to star formation can become comparable to, or even exceed, that of a radio-quiet AGN. While deeper observations hold promise for detecting these radio-quiet signals, the potential contamination from intense star formation activity underscores the challenge of using radio emission alone to unambiguously confirm the AGN nature of LRDs when SFRs are high. This highlights the complementary importance of X-ray constraints, like those presented in this work, which can help disentangle the contributions even in highly star-forming, dusty environments.

A crucial next step is to directly constrain the gas content of LRDs. The presence of a substantial gas reservoir would deepen the central gravitational potential and enhance the rate of dynamical interactions and stellar collisions, potentially influencing the assembly of compact stellar systems and the fueling of central BHs \citep[e.g.][]{Boekholt2018,TAGAWA2020,REINOSO2020, Schleicher2022, Reinoso2023, Solar2025}. Second, measurements of gas columns are necessary to discriminate between different physical interpretations of LRDs: scenarios such as heavily obscured active nuclei or compact starbursts depend sensitively on the assumed hydrogen column density, with obscured AGN models typically requiring $N_{\rm H}\gtrsim 10^{23}$~cm$^{-2}$\citep[e.g.,][]{HICKOX2018}, while star-forming or unobscured models are consistent with much lower values. Therefore, direct determinations of gas mass and column density are essential both for understanding the internal dynamics of these systems and for ruling out competing formation pathways.

\begin{acknowledgements}
ML gratefully acknowledges support from ANID/DOCTORADO BECAS CHILE 72240058. DRGS  gratefully acknowledges the support of the ANID BASAL project FB21003 and the Alexander von Humboldt - Foundation, Bonn, Germany.
MCV acknowledges funding through ANID (Doctorado acuerdo bilateral DAAD/62210038) and DAAD (funding program number 57600326). MCV acknowledges the International Max Planck Research School for Astronomy and Cosmic Physics at the University of Heidelberg (IMPRS-HD).
This material is based upon work supported by Tamkeen under the NYU Abu Dhabi Research Institute grant CASS.
\end{acknowledgements}

\bibliographystyle{aa}
\bibliography{References}

\clearpage

\appendix

\section{MCMC fitting}\label{Apppendix:A}

In Fig. \ref{fig:radiusMassAkins}, we show the results of the best fit of the correlation $M_{\rm gal}-R_{\rm gal}$ for the data sample of \citetalias{AKINS2024}. We assume a linear model in log space with intrinsic scatter $\log_{10}\sigma$,
\begin{equation}
    \log\left(\frac{M_{\rm gal}}{{\rm M}_\odot}\right) = A \log\left(\frac{R_{\rm gal}}{{\rm R}_{\odot}}\right)+B,\label{eq:modelMCMC}
\end{equation}
where $A$ is the slope and $B$ is the intercept parameter. In summary, our statistical model is described by $\theta=(A,B,\log_{10} \sigma)$.

\begin{figure}[!h]
    \centering
\includegraphics[scale=0.9]{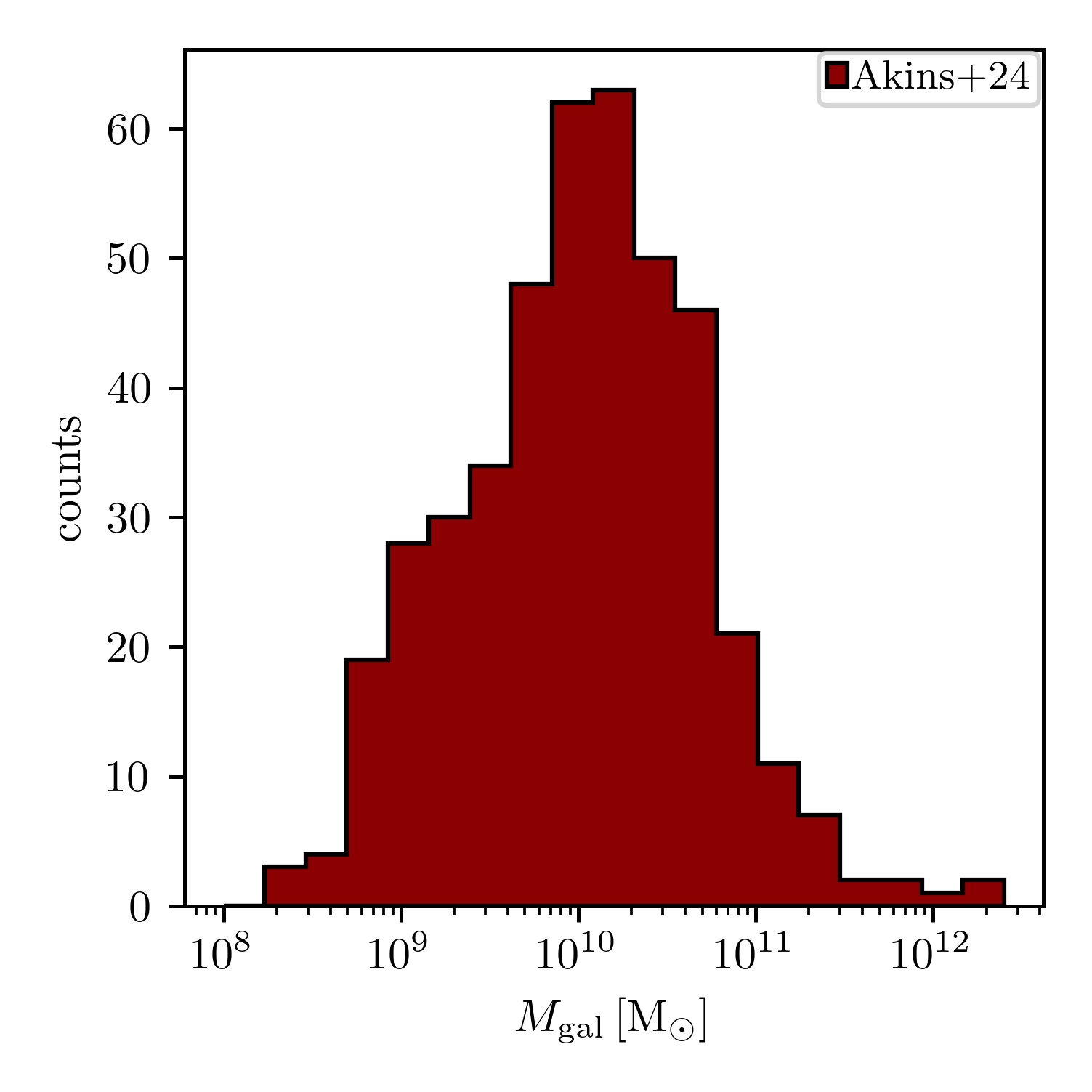}
    \caption{Mass distribution of the LRD galaxies in the \citetalias{AKINS2024} sample.}
    \label{fig:massDistributionAkins}
\end{figure}

The posterior distribution $P(\theta|D)$ is sampled using Bayes' Theorem, i.e, 
\begin{equation}
    P(\theta|D)\propto \mathcal{L}(D|\theta)\cdot \pi(\theta),
\end{equation}
with $\mathcal{L}$ the likelihood function. We assume a Gaussian likelihood where the observed $\log_{10}M_i$ for each galaxy is normally distributed (see Fig. \ref{fig:massDistributionAkins} for the mass distribution) around the model prediction (including the intrinsic scatter) as
\begin{equation}
    \ln{\mathcal{L}(D|\theta)}=-0.5\sum_{i}\left[\frac{(\log_{10}M_i-(A\log_{10}R_i+B))^2}{\sigma^2}-\ln(2\pi\sigma^2) \right],
\end{equation}
where we assume uniform priors over broad ranges to allow the data to dominate the inference. The values of the priors are listed in Table \ref{tab:priorsMCMC}.

\begin{table}[!h]
    \centering
        \caption{The uniform priors $\pi(\theta)$ adopted for the MCMC analysis.}
    \begin{tabular}{cc}
        Parameter & Prior range  \\ \hline\hline
          A &  uniform over $[-3, 3]$\\
        B   & uniform over $[5,15]$ \\ 
       $\log_{10}\sigma$ & uniform over $[-2, 2]$ \\ \hline
    \end{tabular}
    \label{tab:priorsMCMC}
    \tablefoot{The model parameters are $\theta=(A,B,\log_{10}\sigma)$.
    }
\end{table}

The posterior distribution was sampled using the emcee Python package. We used $N_{\rm  walkers} =32$ chains, where the chains were run for a total of 2000 steps. The burn-in phase of 250 steps was discarded to ensure the chains had fully converged to the stable posterior distribution. We verify convergence and adequate sampling by calculating the autocorrelation time $\tau$ for each parameter. The maximum $\tau$ observed was 16.63 steps. Since the chain length (2000 steps) significantly exceeded $50\times \tau$, we conclude that the samples were sufficiently independent and well mixed.

\begin{figure}
    \centering
    \includegraphics[scale=0.5]{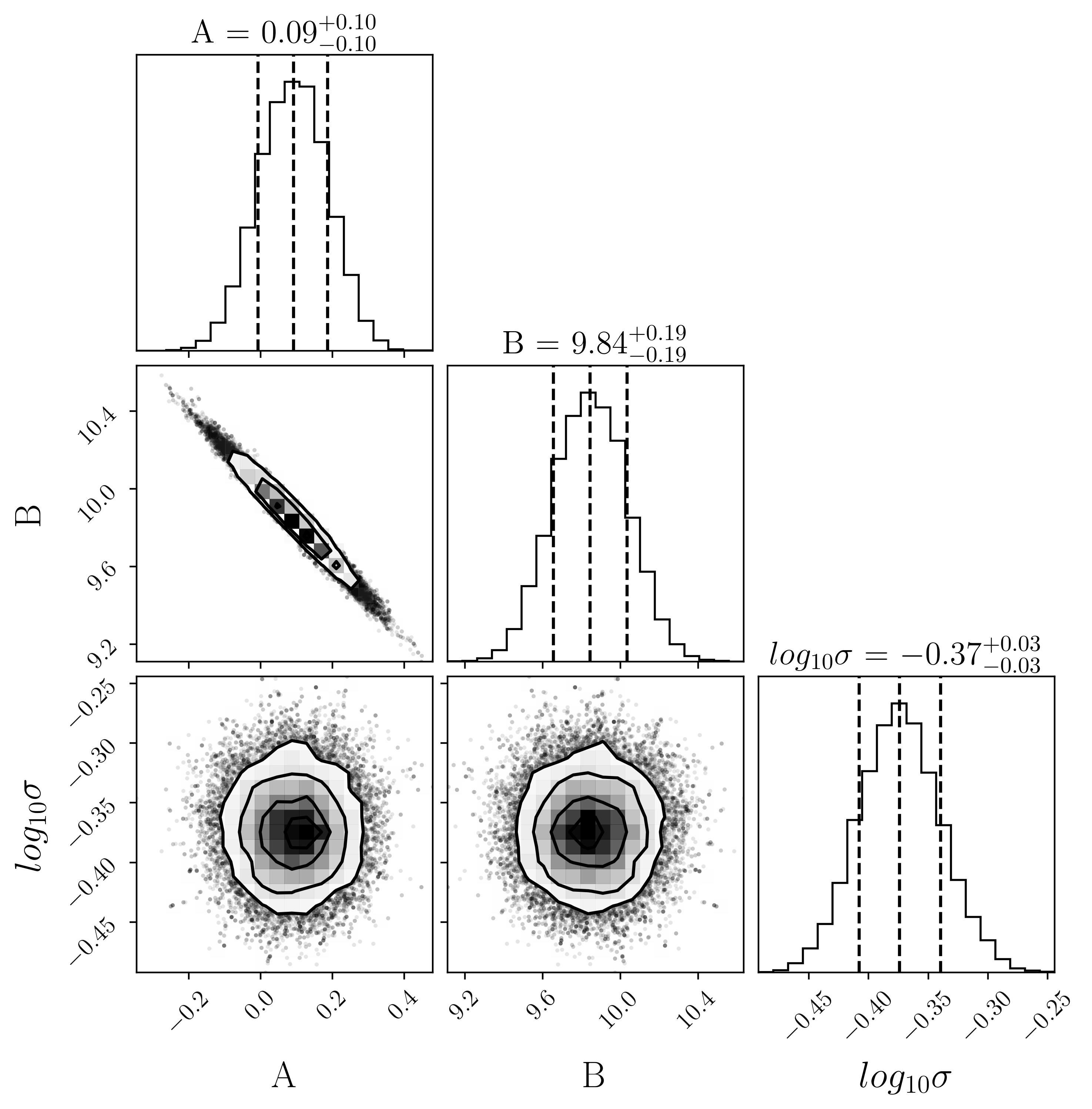}
    \caption{The marginalized one-dimensional (diagonal) and two-dimensional (off-diagonal) posterior probability distributions for the model parameters: slope (A), intercept (B), and log-scatter ($\log_{10}\sigma$). The diagonal histograms show the median (central line) and the $1\sigma$ intervals. A strong negative correlation is evident between the slope (A) and the intercept (B).}
    \label{fig:cornerPlot}
\end{figure}

Our best-fit values are listed in Table \ref{tab:valuesMCMC}. The slope of the assumed correlation ($A=0.09\pm0.10$) is nearly flat.  As the radius increases by a factor of 10, its mass, on average, increases only by a factor of $\approx1.23$. This suggests that galaxies are distributed across the $\log M-\log R$ plane with a mass distribution that is primarily determined by factors other than size. We also emphasize that we consider the upper limits of the radius as almost all of the sources are not spatially resolved. More precise measurements of the radius and including more free parameters  might change this analysis.

\begin{table}[!h]
    \centering
     \caption{Best fit parameters from MCMC analysis.}
    \label{tab:valuesMCMC}
    \begin{tabular}{cc}
        Parameter & Best fit  \\ \hline\hline
          A & $+0.09^{+0.10}_{-0.10}$\\
        B   & $+9.84^{+0.19}_{-0.19}$ \\ 
       $\log_{10}\sigma$ & $-0.37^{+0.03}_{-0.03}$ \\ \hline
    \end{tabular}
    \tablefoot{ We show the results of the best values plus the $1\sigma$ errors.}
\end{table}

\begin{figure}[!h]
    \centering
\includegraphics[scale=0.95]{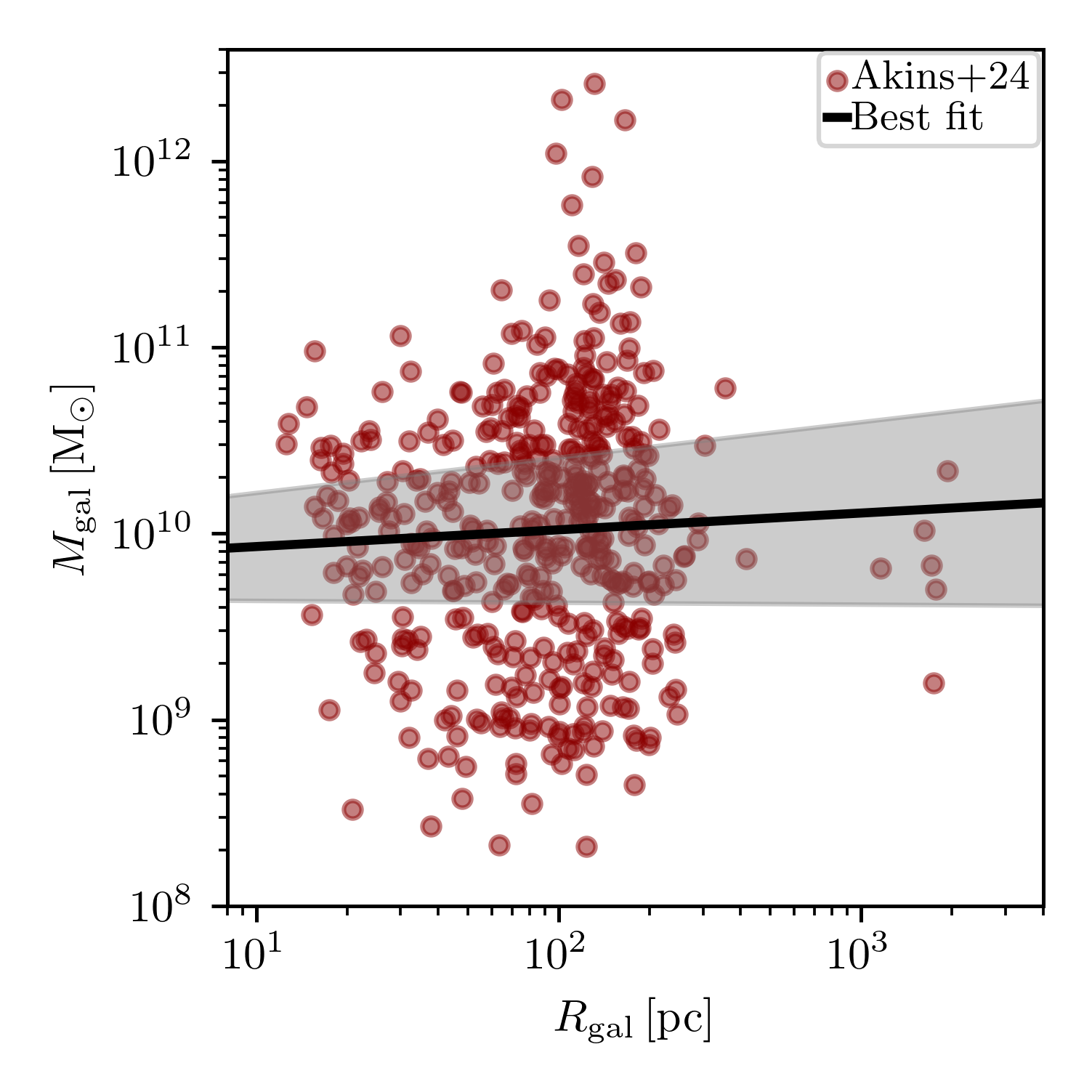}
    \caption{Data and model comparison of  LRD galaxies in the \citetalias{AKINS2024} sample. The dashed black line shows the mean value of the model introduced in Eq. \ref{eq:modelMCMC}, while the shadow corresponds to the $1\sigma$ confidence interval.}
    \label{fig:radiusMassAkins}
\end{figure}

\newpage

\section{Samples comparison}
\label{sec:appendix_comparison}

In this Appendix, we compare the properties of the \citetalias{AKINS2024} sample, which we use to define the structural parameters of Little Red Dots (LRDs), and the subsample of \citetalias{SACCHI2025} used for the X-ray analysis in this work.

\begin{figure}[!h]
    \centering
    \includegraphics[width=0.40\textwidth]{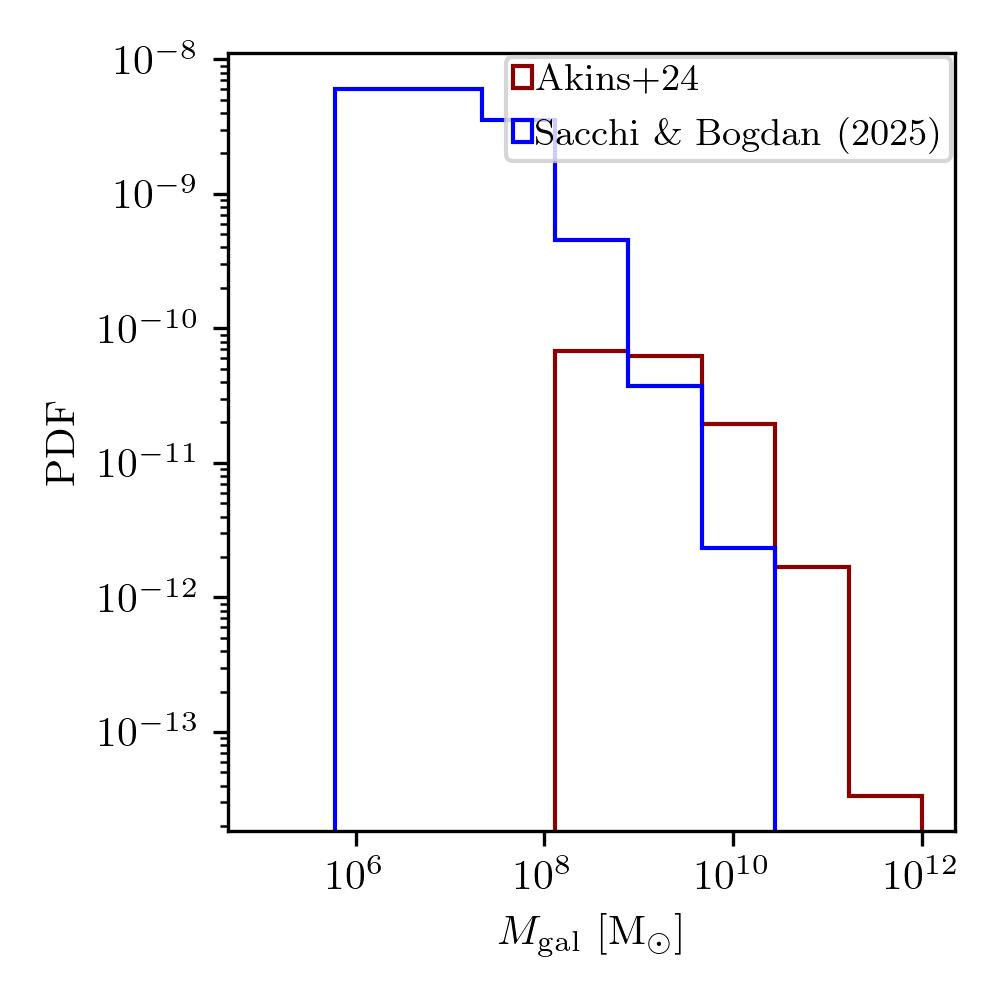}
    \caption{Comparison of the stellar mass distributions between the \citetalias{AKINS2024} sample (red) and the \citetalias{SACCHI2025} subsample (blue). While the \citetalias{SACCHI2025} sample extends to lower masses, the X-ray stacking signal is dominated by the massive end of the population, justifying the use of structural parameters derived from the \citetalias{AKINS2024} sample.}
    \label{fig:massDistribution}
\end{figure}

In Fig. \ref{fig:massDistribution}, we compare the stellar mass distributions of the two samples. The \citetalias{AKINS2024} sample (red) is peaked at high masses ($M_{\rm gal} \sim 10^{10} - 10^{11}$\,M$_\odot$), while the \citetalias{SACCHI2025} subsample (blue) exhibits an extended tail of low-mass objects reaching down to $10^6$\,M$_\odot$. We argue that the inclusion of lower-mass galaxies in the \citetalias{SACCHI2025} sample does not introduce a bias in our modeling. This is due to the inherent nature of X-ray stacking, where the aggregate signal is flux-weighted. Since the X-ray luminosity scales with stellar mass \citep[$L_{X}\propto M_{\rm gal}$, e.g.,][]{ANDERSON2015}, the stacked signal is dominated by the most massive galaxies in the sample. A single galaxy with a mass of $10^{10}$\,M$_\odot$ contributes a flux comparable to thousands of $10^6$\,M$_\odot$ systems. Therefore, the structural parameters of the massive end of the population (represented by the \citetalias{AKINS2024} sample) are the relevant priors for modeling the sources responsible for the potential X-ray emission.

\begin{figure}[!h]
    \centering
    \includegraphics[width=0.40\textwidth]{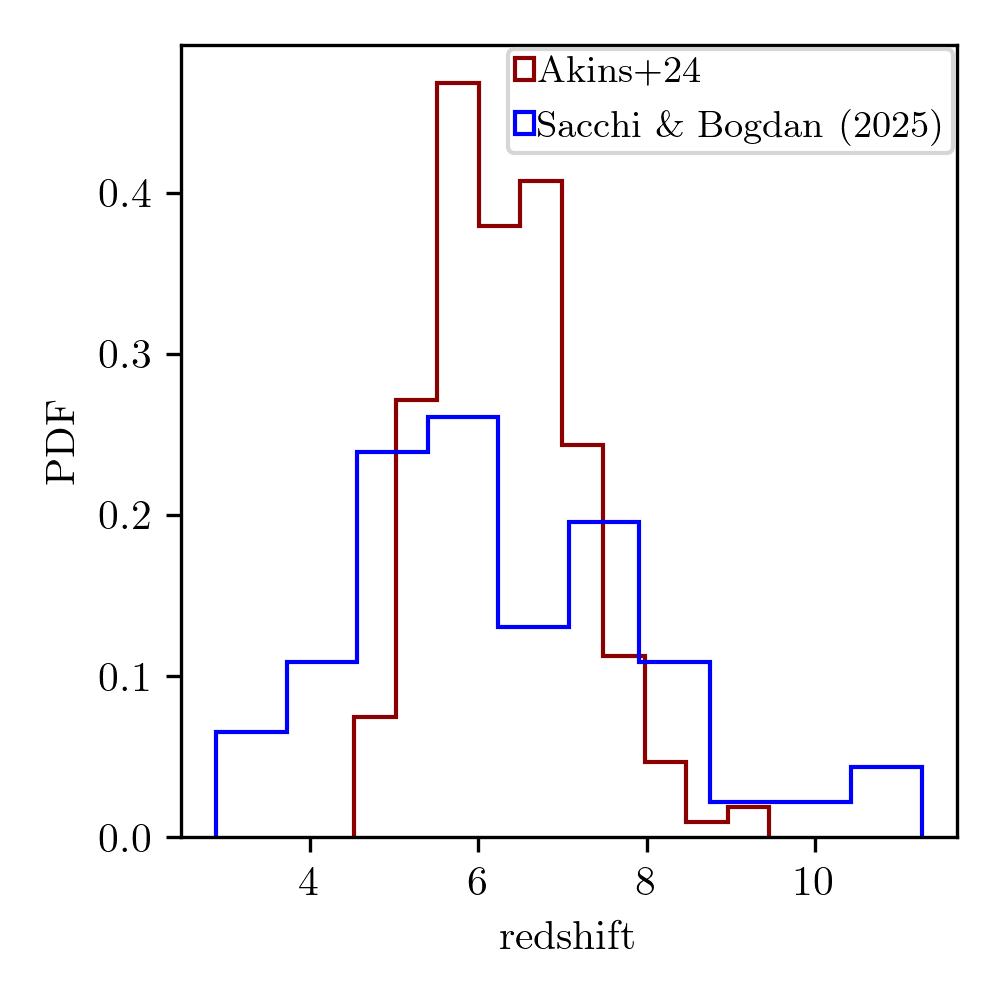}
    \caption{Comparison of the redshift distributions. The samples overlap significantly in the $z=5-8$ range. For the high-redshift tail ($z>8$) in the \citetalias{SACCHI2025} sample, our use of radii from the lower-redshift \citetalias{AKINS2024} sample yields conservative estimates for stellar density and collisional efficiency.}
    \label{fig:redshiftDistribution}
\end{figure}

In Fig. \ref{fig:redshiftDistribution}, we show the redshift distributions. Both samples exhibit significant overlap in the $z=5-8$ range, which dominates the stacking signal. However, the \citetalias{SACCHI2025} sample extends to both lower ($z\sim3$) and higher ($z\sim11$) redshifts. We assume that the structural properties of LRDs remain comparable across this redshift window. For the high-redshift tail ($z > 8$), this is a conservative assumption. Galaxies at higher redshifts are theoretically and observationally expected to be more compact at fixed stellar mass \citep[e.g.,][]{SHIBUYA2015, ALLEN2025}. By applying the radius distribution from the slightly lower-redshift sample of \citetalias{AKINS2024}, we likely overestimate the physical sizes of these highest-redshift sources. In our collision-based model, larger radii imply lower stellar densities and, consequently, lower BH formation efficiencies. Thus, our approach likely underestimates the BH mass density at the high-redshift end, further strengthening our conclusions that collisional formation is a viable channel.

\section{Effect of accretion in final black hole masses}\label{appendix:Accretion}

\begin{figure}[!h]
    \centering
\includegraphics{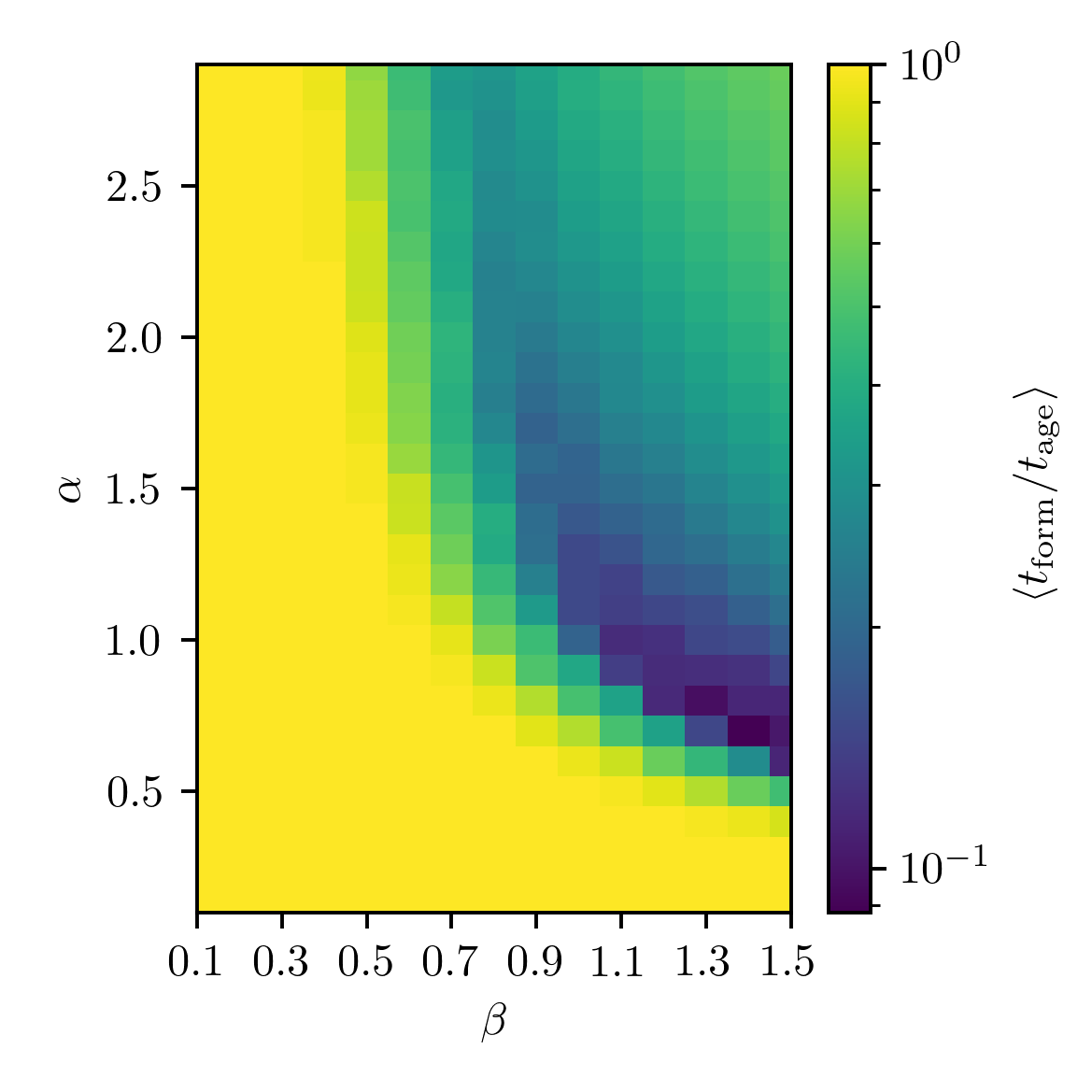}
    \caption{The colorbar shows the average ratio between the BH formation time ($t_{\rm form}$) and the total age of the galaxy ($t_{\rm age}$) for each ($\alpha$, $\beta$) model. The y-axis shows the star formation parameter $\alpha$, and the x-axis shows the mass-radius relation exponent $\beta$. Dark purple regions indicate very early seed formation ($t_{\rm form} \approx 0.1\,t_{\rm age}$), while yellow regions indicate very late formation ($t_{\rm form} \approx t_{\rm age}$). }
    \label{fig:fractionTime}
\end{figure}

\begin{figure}[h!]
    \centering
\includegraphics{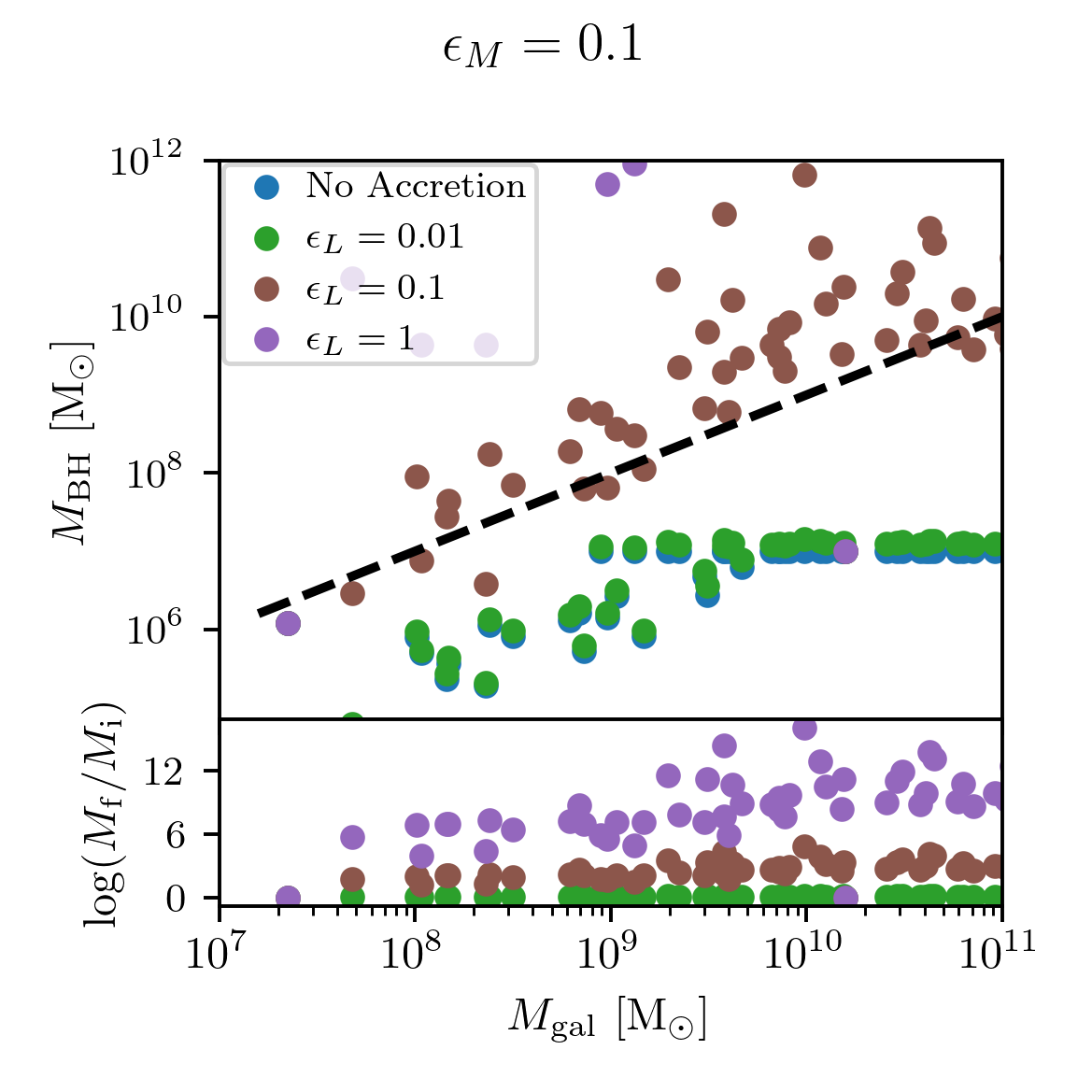}
\includegraphics{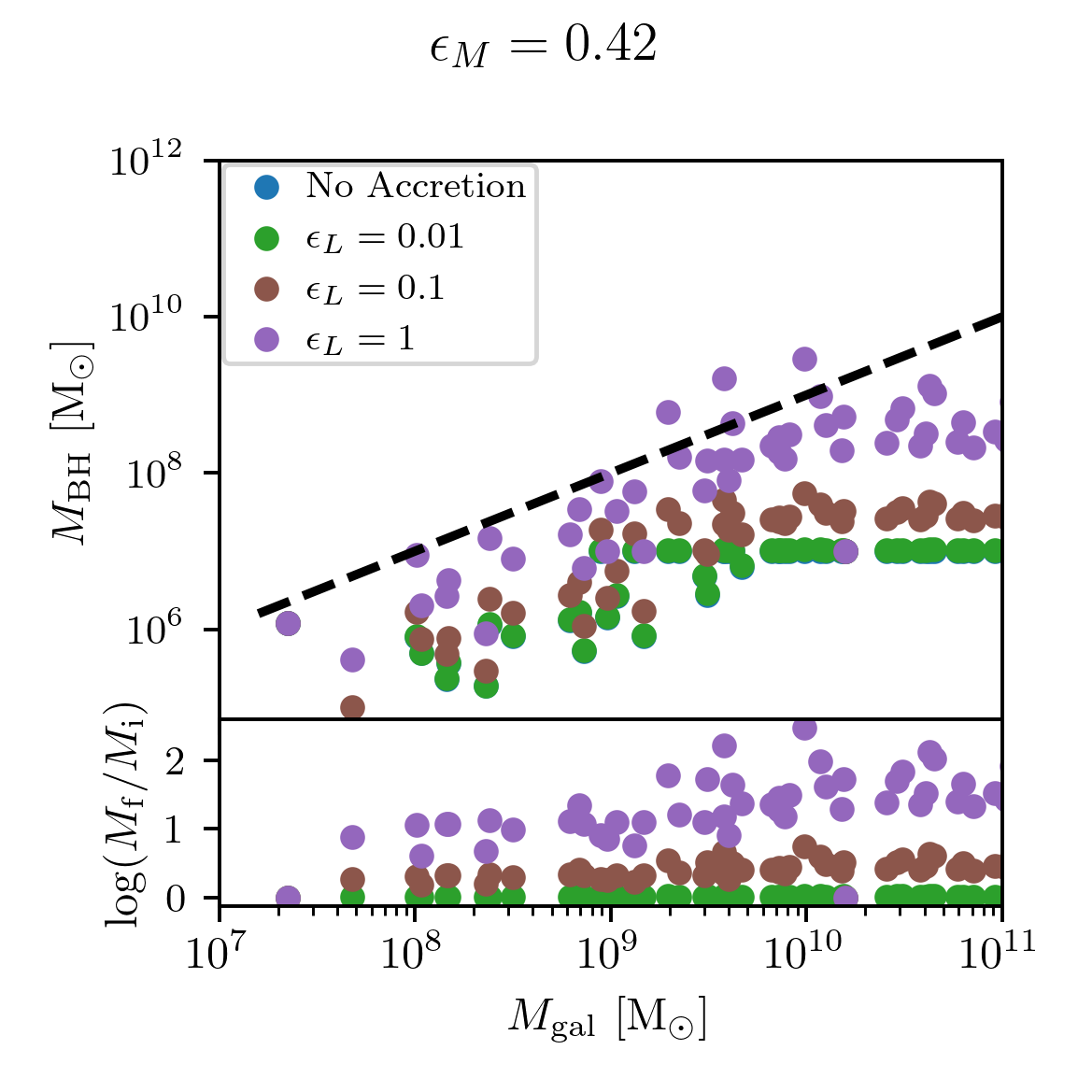}
    \caption{Top panel: BH mass versus host galaxy mass for a fiducial model with $\alpha=0.7$ and $\beta=1.4$ assuming $\epsilon_M=0.1$. Blue dots show the initial seed masses (No Accretion). Green, brown and purple dots show the final mass after accretion with efficiencies $\epsilon_L=0.01$, $\epsilon_L=0.1$  and $\epsilon_L=1$, respectively. The dashed line shows the $M_{\rm BH}/M_{\rm gal} = 0.1 $ for reference. We also show the logarithmic ratio of the final (after accretion) mass to the initial mass for the models, quantifying the total BH growth. 
    Bottom panel: Same as top panel but here we set $\epsilon_M=0.42$ assuming that the BH is efficiently radiating.}
    \label{fig:Rates}
\end{figure}

Fig.~\ref{fig:fractionTime}, we map the average BH formation time ($t_{\rm form}$) as a fraction of the galaxy age ($t_{\rm age}$) across our parameter space of $\alpha$ and $\beta$. First, we note that in models with $\beta\sim0.6$, as in typical spiral galaxies in the Local Universe, the seed BH tends to form rather late. In this case, not much time is available for subsequent growth. In this case, the accretion rate cannot be strongly constrained from the comparison with the observed BH masses, as the time window for accretion is relatively short. We nonetheless also report here about models where the seed BH forms at earlier stages, requiring more extreme assumptions about the $\beta$ parameter.  The colorbar, $\langle t_{\rm form}/t_{\rm age} \rangle$, reveals a strong trend in which models with high $\beta$ ($\gtrsim 1.2$) and low-to-intermediate $\alpha$ ($\sim0.5-1.0$) produce BH seeds extremely early. In these extreme scenarios (dark purple region), seeds form, on average, when the galaxy is only about $10\%$ of its current age ($t_{\rm form} \approx 0.1\,t_{\rm age}$). We investigate the implications of this early formation in Fig.~\ref{fig:Rates}. We select a representative model from this extreme, early-forming regime: $\alpha=0.7$ and $\beta=1.4$, and we calculate the final BH mass after a period of accretion.

 We estimate the final mass $M_f$ at a time t using the expression from \cite{SHAPIRO2005}, which describes exponential growth:\begin{equation}M_f(t) = \exp\left(\epsilon_L\frac{(1-\epsilon_M)}{\epsilon_M}\frac{(t-t_i)}{\tau} \right) M_i(t_i),\label{eq:accretion}\end{equation} where $t_i$ is the formation time of the BH seed (equivalent to $t_{\rm form}$ from Figure 1), $M_i(t_i)$ is the initial seed mass, and $t$ is the final time (corresponding to the age of the galaxy, $t_{\rm age}$). The parameter $\epsilon_L=L/L_{\rm Edd}$ is the accretion luminosity as a fraction of the Eddington luminosity. The radiative efficiency, $\epsilon_M$, is the fraction of accreted rest-mass energy converted to light (typically $\sim0.1$). The characteristic accretion timescale is $\tau=Mc^2/L_{\rm Edd}$, where $M$ is the BH mass and  $c$ is the speed of light. Using this formalism, we analyze the results in Fig.~\ref{fig:Rates}. Here, we select a representative model from the extreme, early-forming regime identified in Fig.~\ref{fig:Rates}, specifically, the $\alpha=0.7$ and $\beta=1.4$ model. We explore the impact of BH radiative efficiency ($\epsilon_M$) and Eddington ratio ($\epsilon_L$) on the final BH mass ($M_{\rm BH}$). The two panels contrast a low-efficiency model,  observed in Quasars \citep[e.g., ][]{SOLTAN1982} with a typical value $\epsilon_M \sim 0.1$ (top panel)
 with a high-efficiency model $\epsilon_M = 0.42$ (bottom panel) which is the upper limit for a thin accretion disk \citep{SHAKURA1973}, both applied to the same initial seed population (blue points).
For the low radiative efficiency ($\epsilon_M = 0.1$), $90\%$ of the accreted mass is retained by the BH, leading to rapid and efficient mass growth. As a result, both Eddington-limited ($\epsilon_L = 1$, purple) and moderate sub-Eddington ($\epsilon_L = 0.1$, brown) accretion rates produce final BH masses that are unphysically large. The $\epsilon_L = 1$ case, in particular, results in growth factors of $\log(M_f/M_i) \sim 6-12$, placing the final masses many orders of magnitude above the maximum ratio observed $M_{\rm BH}/M_{\rm gal}=0.1$ (dashed line). Even the $\epsilon_L = 0.1$ rate produces growth factors of $\log(M_f/M_i) \sim 1-5$, systematically overshooting the relation. Only a very low accretion rate ($\epsilon_L = 0.01$, green) avoids this, but it produces negligible growth. Conversely, the bottom panel assumes a high radiative efficiency of $\epsilon_M = 0.42$, the theoretical maximum corresponding to a maximally spinning Kerr BH. In this scenario, only $58\%$ of the accreted mass is retained, leading to a much more gradual and self-regulated growth. Under this assumption, the model requires a sub-Eddington accretion scenario. We find that a moderate sub-Eddington rate of $\epsilon_L = 0.1$ (brown points) is a physically plausible scenario. It produces a typical growth factor of $\sim 10$. Furthermore, a low sub-Eddington rate ($\epsilon_L = 0.01$, green) is also entirely viable, representing a quiescent population of BHs that have undergone minimal growth since their formation. By comparing these two physically-motivated scenarios, we conclude that sub-Eddington accretion is not only a plausible mechanism but is in fact required to reconcile the initial seed masses with the present-day BH population. A model combining high radiative efficiencies (consistent with spinning BHs) and moderate sub-Eddington accretion rates ($\epsilon_L \approx 0.1$) provides an excellent framework for explaining the co-evolution of BHs and their host galaxies.

We emphasize that the conclusion above—that moderate sub-Eddington accretion is sufficient—relies on the adoption of a collision-favorable evolutionary track ($\alpha=0.7, \beta=1.4$). In this regime, the BH seed forms with a significant fraction of the final mass. In contrast, for 'standard' galaxy evolution scenarios (e.g., $\beta \approx 0.6$, typical of local spirals), the collisional seeds are significantly smaller ($M_{\rm seed} \ll M_{\rm final}$). Consequently, these scenarios require substantial mass growth via gas accretion to match the observed LRD BH masses. This implies a need for higher time-averaged accretion rates (higher $k$) or duty cycles approaching unity. As shown in our main analysis (Fig. \ref{fig:bhMasses}), these accretion-dominated scenarios would result in X-ray fluxes that exceed the stacking upper limits unless extreme obscuration is invoked. Thus, the X-ray non-detections indirectly favor the 'heavy seed' evolutionary paths where the duty cycle can remain low.

\section{High accretion in IR black hole derived masses}\label{appendix:D}

\begin{figure}[h!]
    \centering
    \includegraphics{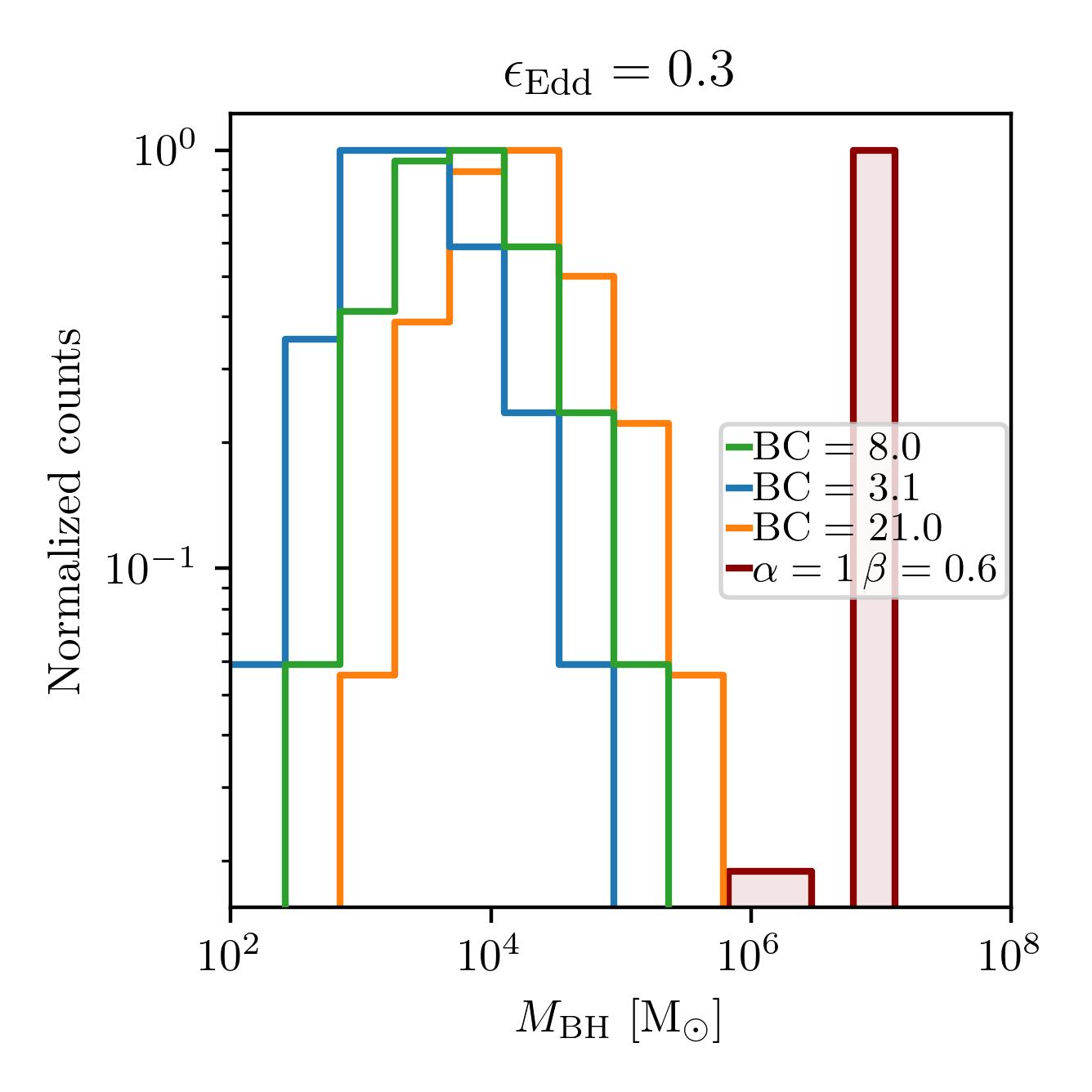}
  \caption{Distribution of BH masses inferred with the luminosities obtained from the $m_{F444W}$ magnitude, compared to our reference model $\alpha=1$ and $\beta=0.6$ (filled red histogram). Top panel assumes an Eddington fraction $\epsilon_{\rm Edd}=10^{-3}$. Bottom panel assumes $\epsilon_{\rm Edd } =10^{-4}$. The colored lines show the mass distributions for the standard bolometric correction (BC=8.0, green), the minimum (BC=3.1, blue), and the maximum (BC=21.0, orange).}    \label{fig:highAccretion}
\end{figure}

In Fig. \ref{fig:highAccretion} we show the comparison between the distribution of BH masses derived from the $mF444W$ magnitude assuming a high Eddington ratio $\epsilon_{\rm Edd}=0.3$. 

In this case, our collision-based scenario would tend to over predict the BH masses estimated from the IR observations. For a high bolometric correction (BC=21, orange distribution), the maximum BH masses estimated are closer to the ones predicted by our collisional model adopting $\alpha=1.0$ and $\beta=0.6$.

\end{document}